\documentclass[preprintnumbers,amsmath,amssymb,floatfix,11pt,prd,onecolumn,superscriptaddress,nofootinbib]{revtex4}
\usepackage{diagbox}
\usepackage{latexsym}
\usepackage{epsfig}
\usepackage{epstopdf,float}
\usepackage{graphicx}
\usepackage{amssymb}
\usepackage{amsmath}
\usepackage{dcolumn}
\usepackage{bm}
\usepackage{color}
\usepackage{comment}
\usepackage[shortlabels]{enumitem}
\usepackage{subfigure}
\usepackage{multirow}
\usepackage{xcolor}
\begin{document}
\title{\bf Kerr-like Black Hole Surrounded by Cold Dark Matter Halo: The Shadow Images and EHT Constraints}
\author{Xiao-Xiong Zeng}
\altaffiliation{xxzengphysics@163.com}\affiliation{State Key
Laboratory of Mountain Bridge and Tunnel Engineering, Chongqing
Jiaotong University, Chongqing $400074$,
China}\affiliation{Department of Mechanics, Chongqing Jiaotong
University, Chongqing $400074$, China}
\author{Chen-Yu Yang}
\altaffiliation{chenyu\_yang2024@163.com}\affiliation{State Key
Laboratory of Mountain Bridge and Tunnel Engineering, Chongqing
Jiaotong University, Chongqing $400074$,
China}\affiliation{Department of Mechanics, Chongqing Jiaotong
University, Chongqing $400074$, China}
\author{M. Israr Aslam}
\altaffiliation{mrisraraslam@gmail.com}\affiliation{Department of
Mathematics, COMSATS University Islamabad, Lahore Campus,
Lahore-$54000$ Pakistan.}
\author{Rabia Saleem}
\altaffiliation{rabiasaleem@cuilahore.edu.pk}\affiliation{Department
of Mathematics, COMSATS University Islamabad, Lahore Campus,
Lahore-$54000$ Pakistan.}
\author{Sadia Aslam}
\altaffiliation{sadiaaslam6466@gmail.com}\affiliation{Department of
Mathematics, COMSATS University Islamabad, Lahore Campus,
Lahore-$54000$ Pakistan.}

\begin{abstract}
Here we provide shadow images of a Kerr-like black hole (BH) in cold dark matter (CDM) halo illuminated with a celestial light source and a thin accretion disk. The impact of spin parameter $a$, critical density $\rho_{c}$ and the scale radius $R_{s}$ on the observed images of BHs is carefully addressed. The results indicate that as $a$ increases, the circular orbits are shifted rightwards, while the larger values of both $\rho_{c}$ and $R_{s}$ are the cause to enhance the radius of circular orbits of the BH shadow. In the case of the celestial light source, the impact of $\rho_{c}$ on shadow distortion is negligible, but this influence is relatively smaller and becomes appreciable when the parameter $R_{s}$ has larger values. Next, we discuss the intensity and the size of the inner shadow, which are gradually increasing with the increase of both $\rho_{c}$ and $R_{s}$. On the other hand, in the case of retrograde flow, the intensity of the shadow images significantly decreases, and a crescent moon emerges on the upper right side of the screen. Subsequently, the distinctive features of red-shift factors for direct and lensed images with prograde and retrograde flows are discussed. The outcomes indicate that the distribution of red-shift factors and the optical appearance are closely related to the behaviour of accreting flow as well as with relevant parameters. Using the recent observational data of EHT, we found that the shadow’s angular diameter of Sgr $A^{\ast}$ provides the best-fit parameter constraints as compared to M$87^{\ast}$.
\end{abstract}
\date{\today}
\maketitle

\section{Introduction}

During the last decades, significant development has been made in the field of BH, especially probing of their shadows and their related consequences. The modern development in the field of gravitational physics has opened a new window in the investigation of electromagnetic radiation to explore the intricate properties of compact objects, including BHs. In $2016$, the detection of the gravitational waves GW$150914$ event by the LIGO-Virgo collaboration, produced by the merger of two BHs having masses $29$ and $36$ times that of the Sun, respectively. This provided strong evidence for the existence of these mysterious entities \cite{sd1,sd2}. Importantly, the earth-shaking discoveries of the Event Horizon Telescope (EHT), which captured the first electromagnetic radiation emitted by the super-heated plasma around the supermassive BH at the heart of the M$87$ and Milky Way galaxies \cite{sd3,sd4}, and the gravity equipment with the discovery of infrared flares in the surroundings of galactic center \cite{sd5}. These novel findings not only confirm the theoretical predictions of general relativity (GR) but also furnish significant data for investigating the intricate properties of BHs and the nature of extreme cosmic environments, opening a new era in BH astronomical research.

An astrophysical BH maintains a stable space-time structure but can be lit by external sources of luminous accretion material, resulting in an extensive range of shapes and colours. If light from an accretion material approaches a BH, the strong gravitational field bends it towards the singularity. This allows it to discuss the optical signatures of the BH from the accretion flow. The complex structure of BH shadow is accurately explained by the fundamental mechanism of photon orbits and space-time geometry around the BH. From the images of the EHT, it has been shown that there is a dark interior region in the centre, which is the so-called BH shadow, and the region outside of which is a compact asymmetric ring, known as the bright photon ring. It is clearly shown by the EHT results that the diameter of the photon ring of Sagittarius (Sgr $A^{\ast}$) is aligned with the radius of the shadow’s critical curve as predicted in GR within the domain of $10\%$ nicely \cite{sd6,sd7,sd8,sd9,sd10}. The investigation of BH shadow has a long history since the early days of GR. In $1960$, initially, the theoretical framework for the explanation of the Schwarzschild BH shadow is discussed in \cite{sd11} and provides the formula to measure the angular radius of the BH shadow. After that, Bardeen analysed the Kerr BH shadow and taking advantage of the separability of the null geodesic equations, constructed a formalism for obtaining the boundary of the shadow \cite{sd12}. Despite its significance, these advances viewed the BH shadow as a theoretical phenomenon which is unlikely to be observed experimentally. In \cite{sd13}, the possibility of viewing the BH shadow of our Milky Way galaxy was proposed along with the requisite experimental conditions.

The investigation of observational features of BHs is an important clue to explore the nature of BHs. Consequently, among the pool of frameworks and proposals to extend GR, some extensive investigations were carried out on the BH shadows along with different contexts of BH physics such as regular BHs \cite{sd14,sd15}, non-commutative BHs \cite{sd16,sd17}, BHs in modified theories of gravity \cite{sd18,sd19,sd20} and so on. Moreover, substantial developments have been made in the investigation of BH shadows and Einstein rings with the help of a wave optics setup \cite{sd21,sd22,sd23,is1}. The acquisition of BH images is a key milestone in BH research and has enormous scientific implications. These images provide insights into the accretion, radiation, and jet phenomena close to BHs as well as their space-time features. It is well-known that in a realistic astronomical situation, a supermassive BH is surrounded by a huge amount of high temperature radiative plasma, resulting in a bright accretion disk. The thermally synchronised electrons act as the paramount luminous source of light in BH image processing. Moreover, EHT results indicate that the magnetic plasma around the accretion disk is constrained within an accurate measurement of the predictions obtained from the models of general relativistic magnetohydrodynamic simulation. Subsequently, the complex structure of the bright regions is intricately related to the accretion disk characteristics, which are affected by the particular accretion disk model and the underlying physical framework.

In various analyses on BH shadows, scientists investigated the influence of different accretion models on the BH images. To see a comprehensive optical appearance of the BH shadows and the distortion of light around them, considering a four colours celestial light source model, the authors in \cite{sd24} numerically investigated the Einstein rings, images of BH shadows as well as the space-dragging effect caused by rotation. Based on this method, Zhong et al. \cite{sd25} discussed the shadows of Kerr BH immersed in a uniform electromagnetic field with the help of a backwards ray-tracing numerical procedure. Hioki and Maeda \cite{sd26} investigated the Kerr BH shadow or a Kerr naked singularity through the development of two observable attributes of contour plots. In the framework of modified theories of gravity, the shadows cast by rotating and non-rotating BH were investigated under different values of parameters \cite{sd27,sd28}. The observational appearance of the magnetic field around the Kerr-Melvin BH is investigated in \cite{sd29}, where the authors concluded that the inner shadow and the critical curve can be interpreted to approximate the magnetic field around a BH without degeneration. By employing the backwards ray-tracing method, Yang et al. \cite{sd30} investigated the intricate properties of rotating Ghosh-Kumar BH shadow images in the background of the celestial light source and thin accretion disk. Liu and his collaborators used the Hamiltonian constraint approach to analyse the light rings and shadows of two types of static BHs that preserve general covariance \cite{sd31}. The study used topological methods and the backwards ray-tracing technique to classify the light rings as standard or unstable. In the context of scalar-tensor-vector gravity, the authors in \cite{sd32} indicated that the cosmological constant and scalar-tensor vector gravity parameter led to intriguing phenomena in the BH shadow contours. Through the energetic achievements of researchers, substantial contributions have been made in the theoretical study of BHs shadows and accretion disks through the effective implementations, making significant improvements to astronomical observations \cite{sd33,sd34,sd35,sd36,sd37,sd38,sd39}. Additionally, due to the similarity between BH shadows and boson star observational images, some significant studies on boson star's optical images under different potentials and gravitational modifications have been discussed in \cite{sd40,sd41,sd42}.

The development and evolution of the large-scale structure of our
cosmos is one of the most fundamental and an emerging area of
research in physical cosmology. Dark matter (DM) halos,
gravitationally apprenticed concentrations of DM play a substantial
role in the complex, non-linear processes involved in combining
cosmic formation. These halos serve as the cornerstone on which
luminous matter gathers, affecting the creation of galaxies,
clusters of galaxies, and super-clusters of galaxies. Hence,
comprehending the complex properties and behaviour of DM halos is
crucial for understanding the cosmic structure formation and its
relationship with cosmological models \cite{sd43}. Additionally, in
Newtonian mechanics, the rotational speed of stars is
$v^{2}=\frac{GM}{r}$, interpreting that beyond the edges of
galaxies, the rotational speed of stars decreases with the
augmentation of their distances from the centre of the galaxy. However,
it is noticed that the rotational speed of stars and the Milky Way	
behave nearly constant at distances several times that of the
galactic centre \cite{sd44}. Therefore, scientists have hypothesized
the presence of DM and nowadays, DM has been confirmed through
numerous indirect evidence \cite{sd45}. Consequently, a lot of
research has been done on DM, leading to the proposal of several
models such as the CDM model \cite{sd46}, warm DM model \cite{sd47},
etc. In the realm of cosmological models, CDM makes a substantial
contribution to the overall density of the universe, accounting for
$26\%$ of it \cite{sd48}. According to Navarro-Frenk-White (NFW)
profile, the density distribution of the DM halo in the CDM model is
defined as \cite{sd49,sd50}
\begin{equation}\label{s1}
\rho_{_{\text{NFW}}}=\frac{\rho_{c}}{\frac{r}{R_{s}}(1+\frac{r}{R_{s}})^{2}},
\end{equation}
in which $\rho_{c}$ represents the critical density and $R_{s}$ is
the scale radius of the DM halo.

The relation between BHs and DM is an important topic
among the scientific community. Numerous studies interpreted that
the DM in the vicinity of BHs can produce a spiking phenomenon
\cite{sd51} and the existence of DM facilitates the formation
of BHs \cite{sd52}. It has been observed that supermassive BHs
prevail along with the particles of DM in the centre of
galaxies \cite{sd53,sd54}. In this context, the thermodynamic
properties of a rotating BH in the presence of CDM are investigated
in \cite{sd55}, where the stable BHs are found only for smaller
values of $\rho_{c}$. In \cite{sd56}, authors explored the weak
cosmic censorship conjecture in BHs that are closer to those
existing in the real universe, such that rotating BHs enveloped by
DM. The motion spinning particles around the BH in a DM halo are
investigated in \cite{sd57}, where the authors concluded that the
presence of DM halos can permanently vary the orbital
eccentricity, energy, and the innermost stable circular orbit (ISCO)
parameters of spinning test particles. Generally, the astrophysical
BHs are not isolated compact bodies. Their accretion disk is
involved in the region of the DM halo, which supposedly engulfs the
whole Milky Way galaxy. The radiated photons emitted from
accreting matter are also lensed in the DM halo, potentially
enhancing the resulting optical appearance \cite{sd58}. In this
scenario, this work will focus on the astronomical observable
effects of rotating BHs in the presence of CDM, providing
significant information about the influence of variations in
parameter space and observational angle on the optical signatures of
the BH shadow.

The segments of this paper will show up in this order. In Sec. {\bf
II}, we will briefly define the rotating BH with a CDM halo and the
shadow contours, which represent the impact of the critical density
$\rho_{c}$ and the halo core radius $R_{s}$. In Sec. {\bf III}, we
will discuss the visual properties of shadow images in the
background of a celestial light source. In Sec. {\bf IV}, we will
comprehensively review the accretion disk model and its geometrical
framework with a schematic diagram. Moreover, we will also discuss the
impact of variations of parameters on shadow images cast by a thin
accretion disk model, distribution of red-shift factors, lensing
bands under the prograde and retrograde accretion flow and the
comparison of EHT results. Finally, the last section will be devoted
to concluding remarks.

\section{ROTATING BLACK HOLE in the Presence of Cold Dark Matter}

\begin{figure}[H]
\centering
\includegraphics[width=18cm,height=9cm]{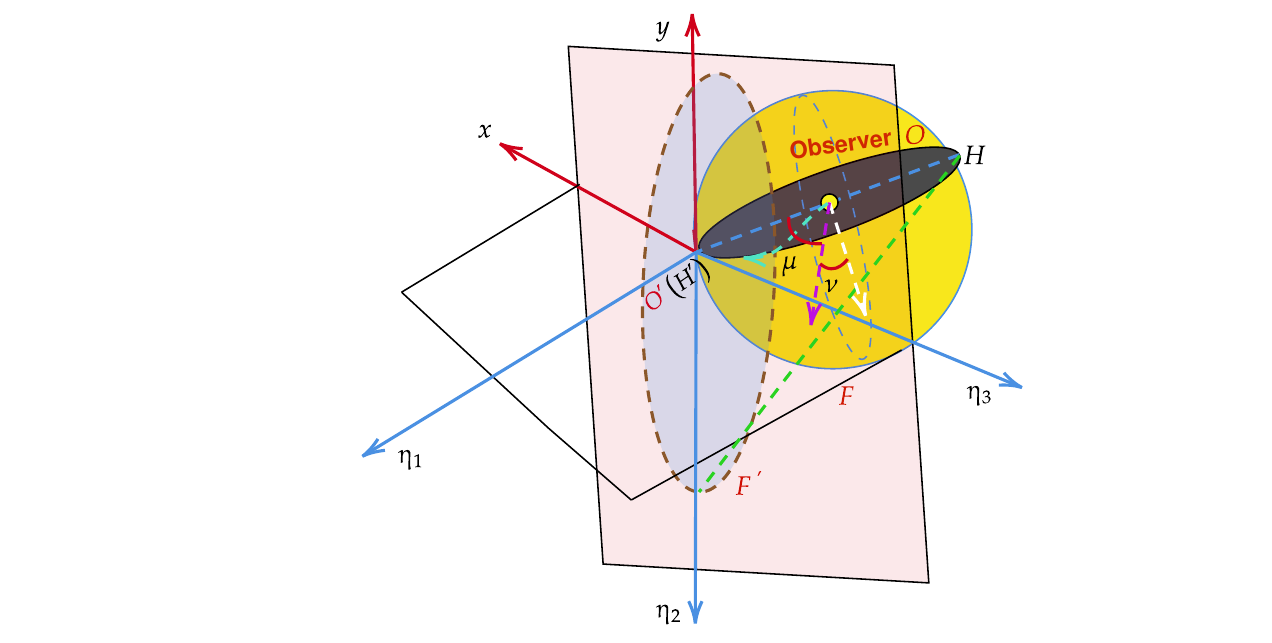}
\caption{The celestial coordinates $(\mu,~\nu)$ are introduced to
imprint each light ray in the observer's frame. Based on
stereographic projection, an appropriate map is illustrated from the
celestial sphere to the screen of our camera.}\label{sf1}
\end{figure}

\begin{figure}[H]\centering
\subfigure[\tiny][]{\label{alb1}\includegraphics[width=4.8cm,height=4.8cm]{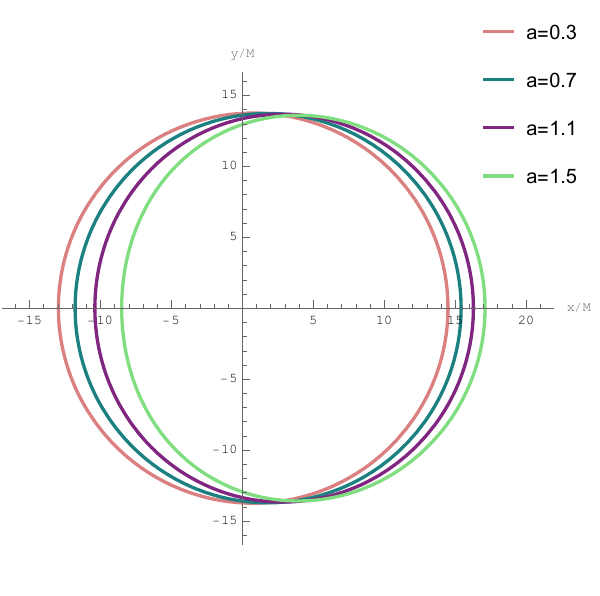}}
\subfigure[\tiny][]{\label{alb2}\includegraphics[width=4.8cm,height=4.8cm]{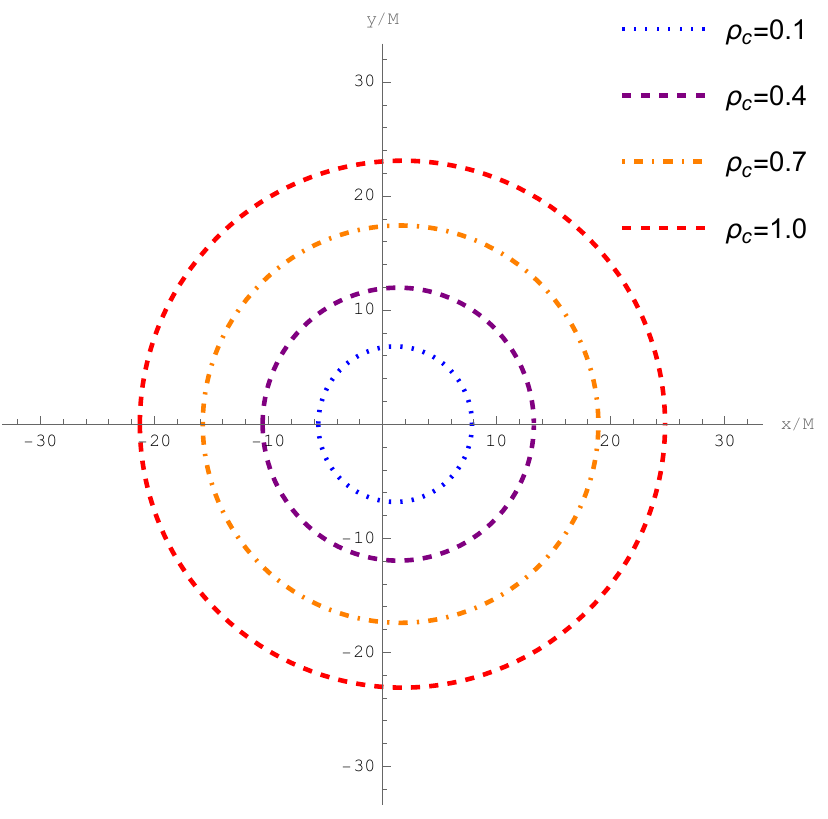}}
\subfigure[\tiny][]{\label{alb2}\includegraphics[width=4.8cm,height=4.8cm]{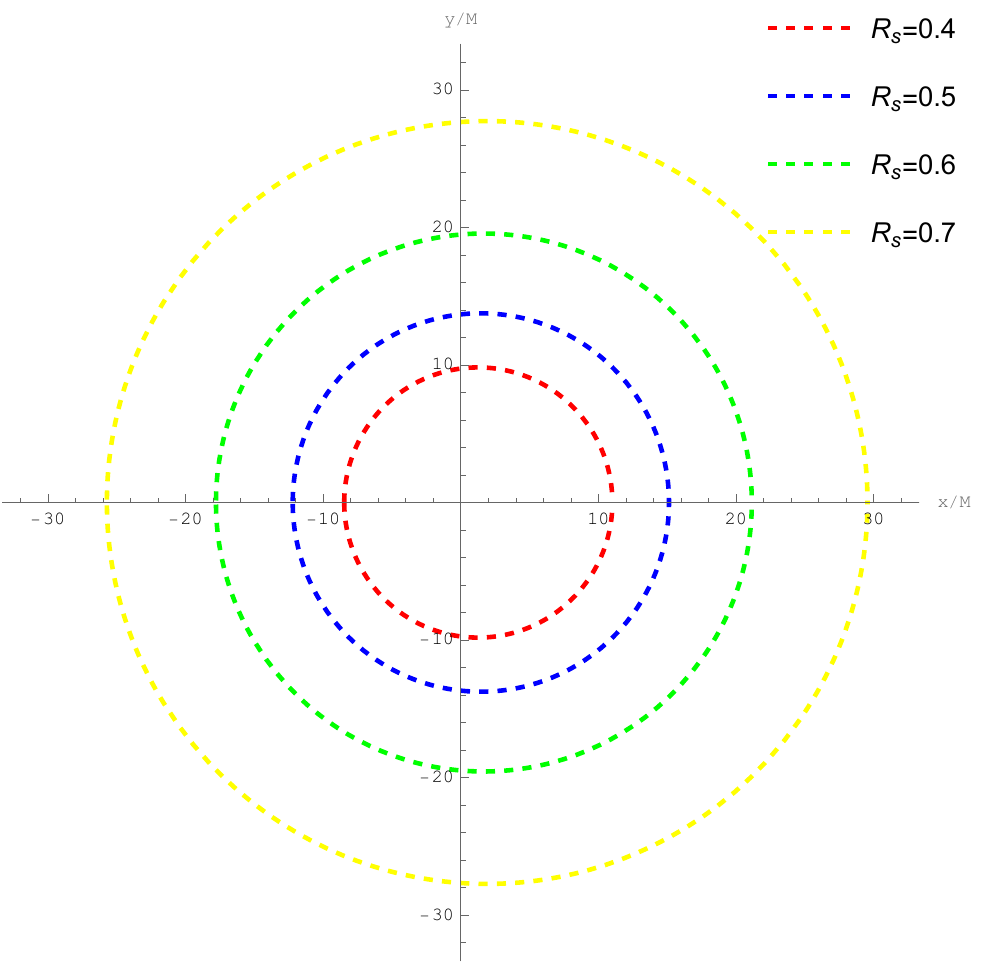}}
\caption{Plots are showing the BH shadows for different values of
$a$, with fixed $R_{s}=0.5=\rho_{c}$ (left panel), for different
values of $\rho_{c}$ with fixed $R_{s}=0.5=a$ (middle panel) and for
different values of $R_{s}$ with fixed $\rho_{c}=0.5=a$ (right
panel). For all cases, we fixed $\theta_{obs}=60^{\circ}$. Further,
the horizontal and vertical axis correspond to $x/M$ and $y/M$,
respectively.}\label{sf2}
\end{figure}

Let us start by introducing the underlying background geometry.
The occurrence of DM halos in galaxies may be defined by many
characteristics such as scale radius and critical density resulting
from particle interactions. When a BH is in the centre of a DM halo,
it can interact with the surrounding DM. The DM BH mechanism can be
accurately characterized by imposing the steady state approximate BH
metric. In the framework of CDM halo as mentioned in Eq. (\ref{s1}),
the space-time metric of a rotating BH is defined as
\cite{sd55,sd59}
\begin{eqnarray}\nonumber
ds^{2}&=&-\bigg(1-\frac{r^{2}+2Mr-r^{2}\big(1+\frac{r}{R_{s}}\big)^{\frac{-8\pi
\rho_{c}R^{3}_{s}}{r}}}{\widehat{\Sigma}^{2}}\bigg)dt^{2}+\bigg(r^{2}+2Mr-r^{2}\big(1+\frac{r}{R_{s}}\big)^{\frac{-8\pi
\rho_{c}R^{3}_{s}}{r}}\bigg)\frac{2a\sin^{2}\theta}{\widehat{\Sigma}^{2}}dtd\phi
\\\label{s2}&+&\bigg((r^{2}+a^{2})^{2}-
\Delta
a^{2}\sin^{2}\theta\bigg)\frac{\sin^{2}\theta}{\widehat{\Sigma}^{2}}d\phi^{2}+\frac{\widehat{\Sigma}^{2}}{\Delta}dr^{2}+\widehat{\Sigma}^{2}d\theta^{2},
\end{eqnarray}
in which
\begin{eqnarray}\nonumber
\Delta&=&a^{2}-2Mr+r^{2}\big(1+\frac{r}{R_{s}}\big)^{\frac{-8\pi
\rho_{c}R^{3}_{s}}{r}},\\\nonumber
\widehat{\Sigma}^{2}&=&r^{2}+a^{2}\cos^{2}\theta,
\end{eqnarray}
where $M$ is the BH's mass and $a$ is rotational parameter. When the
DM critical density $p_{c}$ approaches zero, the metric (\ref{s2})
is reduced to the Kerr BH metric. Now we discuss the motion of photons
around the rotating BH in CDM. Since the photon follows the null
geodesics in a given BH space-time, hence, for space-time
(\ref{s2}), the geodesic motion is governed by the Hamilton-Jacobi
equation, which is given by \cite{sd60}
\begin{eqnarray}\label{s3}
\frac{\partial \mathcal{I}}{\partial
\sigma}=-\frac{1}{2}g^{\xi\zeta}\frac{\partial \mathcal{I}}{\partial
x^{\xi}}\frac{\partial \mathcal{I}}{\partial x^{\zeta}},
\end{eqnarray}
in which $\sigma$ is the affine parameter, $\mathcal{I}$ is the
Jacobi action of the photon. The Jacobi action $\mathcal{I}$ of the
photon can be separated into the following form
\begin{equation}\label{s4}
\mathcal{I}=\frac{1}{2}\gamma^2\sigma-\hat{E}
t+\hat{L}\phi+B_r(r)+B_{\theta}(\theta),
\end{equation}
in which $\gamma=0$ is the photon's mass. The constants
$\hat{E}=-p_{t}$ and $\hat{L}=p_{\phi}$ denote the conserved energy
and conserved angular momentum of the photon in the direction of
rotation axis, respectively. The functions $B_r(r)$ and
$B_{\theta}(\theta)$ are arbitrary functions depend only on the
specified coordinates. Substituting Eq. (\ref{s4}) into Eq.
(\ref{s3}), one can obtain the following equations of motion for the
evolution of the photon
\begin{eqnarray}\nonumber
\widehat{\Sigma}^{2}\frac{dt}{d\sigma}&=&a(\hat{L}-a\hat{E}\sin^{2}\theta)+\frac{r^{2}+a^{2}}{\Delta}(\hat{E}(r^{2}+a^{2})-a\hat{L}),\\\nonumber
\widehat{\Sigma}^{2}\frac{dr}{d\sigma}&=&\pm\sqrt{\mathcal{R}(r)},\\\nonumber
\widehat{\Sigma}^{2}\frac{d\theta}{d\sigma}&=&\pm\sqrt{\Theta(\theta)},\\\label{s5}
\widehat{\Sigma}^{2}\frac{d\phi}{d\sigma}&=&(\hat{L}\csc^{2}\theta-a\hat{E})+\frac{a}{\Delta}(\hat{E}(r^{2}+a^{2})-a\hat{L}),
\end{eqnarray}
with
\begin{eqnarray}\nonumber
\mathcal{R}(r)&=&(\hat{E}(r^{2}+a^{2})-a\hat{L})^{2}-\Delta(\mathcal{J}+(\hat{L}-a\hat{E})^{2}),\\\label{s6}
\Theta(\theta)&=&\mathcal{J}+\big(a^{2}\hat{E}^{2}-\hat{L}^{2}\csc^{2}\theta\big)\cos^{2}\theta,
\end{eqnarray}
where $\mathcal{J}$ is the Carter constant. With the help of these
equations, one can determine the motion of photons around the BH.
Since the photons move in circular orbits, and these orbits together
in all directions, make up a sphere whose radius is represented by
$r_{ps}$. Generally, the condition of photon sphere satisfies the
equation $r=$ constant. Hence, we have $\dot{r}=0=\ddot{r}$ (where
``dot'' is the derivative with respect to affine parameter
$\sigma$), which is equivalent to $\mathcal{R}(r_{ps})=0$ and
$\partial_{r}\mathcal{R}(r_{ps})=0$. Next, we associate the
quantities $\hat{E}$,~ $\hat{L}$ and $\mathcal{J}$ to the impact
parameters close to the BH as
\begin{eqnarray}\label{s6}
\alpha=\frac{\hat{L}}{\hat{E}}, \quad \quad \beta =
\frac{\mathcal{J}}{\hat{E}^2}.
\end{eqnarray}
Now Eq. (\ref{s6}) gives the values of impact parameters as
\begin{eqnarray}\label{s7}
\alpha(r_{ps})&=&\frac{(a^{2}+r^{2}_{ps})\Delta'(r_{ps})-4r_{ps}\Delta(r_{ps})}{a\Delta'(r_{ps})},\\\label{s8}
\beta(r_{ps})&=&\frac{r^{2}_{ps}(-16\Delta(r_{ps})^{2}-r^{2}_{ps}\Delta'(r_{ps})^{2}+8\Delta(r_{ps})(2a^{2}+
r_{ps}\Delta'(r_{ps})))}{a^{2}\Delta'(r_{ps})^{2}},
\end{eqnarray}
where the symbol ($'$) is the derivative with respect to $r$. For an
observer, which is located at infinity, can be defined
as a zero-angular-momentum observer (ZAMO) at coordinates
($t_{obs}=0,~r_{obs},~\theta_{obs},~\phi_{obs}=0$) by assuming the
symmetries in the directions of $t$ and $\phi$. And hence, a locally
orthonormal frame can be developed within the environment of the
observer, which is
\begin{eqnarray}\nonumber
&&\eta_{0}=\eta_{(t)}=\bigg(\sqrt{\frac{g_{\phi\phi}}{g^2_{t\phi}-g_{tt}g_{\phi\phi}}},~0,~0,~-\frac{g_{t\phi}}{g_{\phi\phi}}\sqrt{\frac{g_{\phi\phi}}{g^2_{t\phi}-g_{tt}g_{\phi\phi}}}\bigg),\;\;\;\;\;\;\;\;\;\;\;
\eta_{1}=-\eta_{(r)}=\bigg(0,-\frac{1}{\sqrt{g_{rr}}},~0,~0\bigg)\\\label{s7}&&
\eta_{2}=\eta_{(\theta)}=\bigg(0,~0,~\frac{1}{\sqrt{g_{\theta\theta}}},~0\bigg),\;\;\;\;\;\;\;\;\;\;\;
\eta_{3}=-\eta_{(\phi)}=\bigg(0,~0,~0,-\frac{1}{\sqrt{g_{\phi\phi}}}\bigg),
\end{eqnarray}
where, $\eta_{0}$ represents the time-like vector corresponds to the
observer's $4$-velocity, $\eta_{1}$ indicates the spatial direction
toward the centre of the BH, and $g_{\xi\zeta}$ is the background BH
metric. To see the BH shadow on the observer's screen, an
adequate method based on the pinhole camera model is being
considered \cite{sd61}. This model is simple and
interpret the definite imaging assumptions effectively; however, it has the
limitations due to a narrow field of view. In this scenario, we
closely followed the methodology as defined in \cite{sd62}, which is
well-known as the fisheye lens camera model. Via this
mechanism, in Fig. \textbf{\ref{sf1}}, we show the illumination,
which is useful to see the BH shadow image on the observer's screen,
where the procedure of the stereographic projection technique is
being used.
From Fig. \textbf{\ref{sf1}}, the location of observer is indicated
by the point $O$, whereas the green curve represents the direction
in which the photon traveling along the geodesic at the observer
position. Further, we denote the tangent vector
$\overrightarrow{OF}$ of the null geodesic at the point $O$ in the
$3$-dimensional space. In order to describe the motion of photon
which is seen by the observer, necessitates to consider the
celestial coordinates $(\mu,~\nu)$. Particularly, the development
comprise a $3$-dimensional sphere having centered at $O$ with radius
$OF$. The diameter $HH'$ corresponds to $\eta_{1}$ on the equatorial
plane, the angle between $OH'$ and $\overrightarrow{OF}$ induce the
first celestial coordinate $\mu$ of the optical image. On the
imaging plane, the central point $H'$ lies on the line containing
the diameter $HH'$, which is perpendicular to the equatorial plane,
whereas $\overrightarrow{O'F'}$ indicates the projection of the
vector $\overrightarrow{OF}$ on to the imaging frame. The point
$\mathcal{Q}$ is appointed as the crossing between the line
associating the points $F'$ and $H$ and the $3$-dimensional sphere.
Further, the second celestial coordinate $\nu$ indicates the angle
establish by $O\mathcal{Q}$ and $\eta_{2}$. For a null geodesic,
i.e.,
$S(\sigma)=(t(\sigma),~r(\sigma),~\theta(\sigma),~\phi(\sigma))$,
its tangent vector should be a linear combination of
($\eta_{0},~\eta_{1},~\eta_{2},~\eta_{3}$), which has the following
expression
\begin{equation}\label{s8}
\dot{S}=|\overrightarrow{OF}|(-\chi\eta_{0}+\cos\mu\eta_{1}+\sin\nu\cos\mu\eta_{2}
+\sin\mu\sin\nu\eta_{3}),
\end{equation}
where the negative sign signifies that the tangent vector is
directed in the direction of the past. Since the light ray is free
from photon energy, and hence, one can consider that the energy of
the photon observed in camera's frame is equal to unity, i.e.,
$\hat{E}_{\text{camera}}=1=|\overrightarrow{OF}|.\chi=-\hat{E}(g_{tt})^{-\frac{1}{2}}|_{(r_{O},~\theta_{O})}$.
Additionally, in the frame of ZAMO, the $4$-momentum of photons can
be illustrated as $p_{(\xi)}=p_{\zeta}\eta^{\zeta}_{(\xi)}$, where
the components of $\eta^{\zeta}_{(\xi)}$ are defined in Eq.
(\ref{s7}). The relationship between photon $4$-momentum and the
celestial coordinates $(\mu,~\nu)$ are defined as \cite{sd62}
\begin{eqnarray}\label{s9}
\cos\mu=\frac{p^{(1)}}{p^{(0)}},\quad
\tan\nu=\frac{p^{(3)}}{p^{(2)}}.
\end{eqnarray}
The observer frame can be equipped with a standard Cartesian
coordinate system $(x,~y)$, which developed an accurate align with
celestial coordinates as
\begin{eqnarray}\label{s10}
x(r_{ps})=-2\tan\frac{\mu}{2}\sin\nu, \quad
y(r_{ps})=-2\tan\frac{\mu}{2}\cos\nu.
\end{eqnarray}
\begin{figure}[H]
\begin{center}
\subfigure[\tiny][$R_{s}=\rho_{c}=0.5$]{\label{a1}\includegraphics[width=5cm,height=5cm]{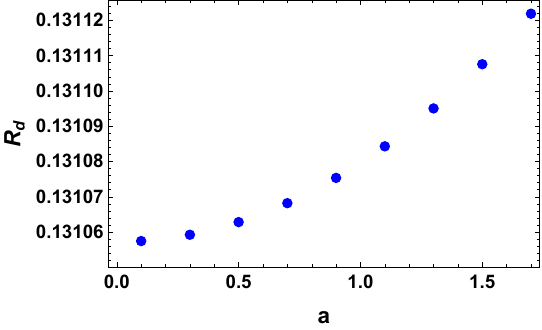}}
\subfigure[\tiny][$R_{s}=a=0.5$]{\label{b1}\includegraphics[width=5cm,height=5cm]{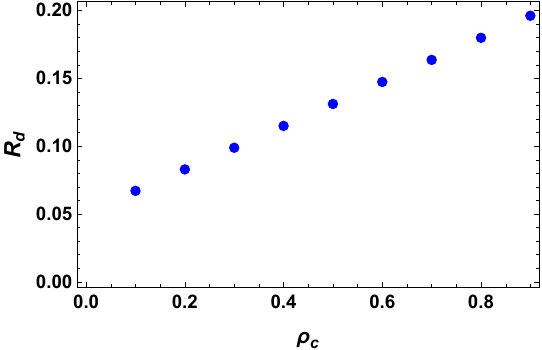}}
\subfigure[\tiny][$\rho_{c}=a=0.5$]{\label{c1}\includegraphics[width=5cm,height=5cm]{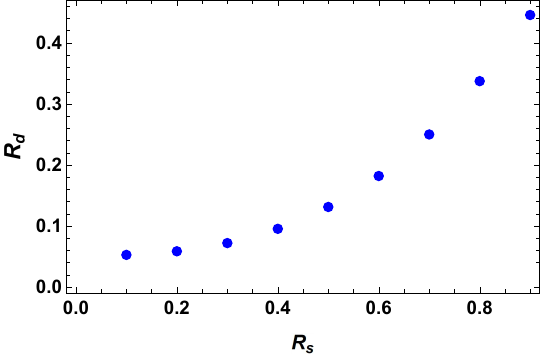}}
\subfigure[\tiny][$R_{s}=\rho_{c}=0.5$]{\label{d1}\includegraphics[width=5cm,height=5cm]{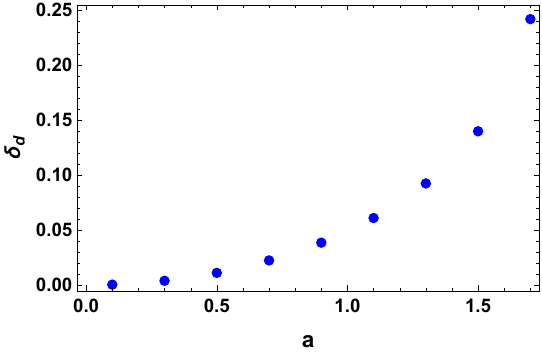}}
\subfigure[\tiny][$R_{s}=a=0.5$]{\label{a2}\includegraphics[width=5cm,height=5cm]{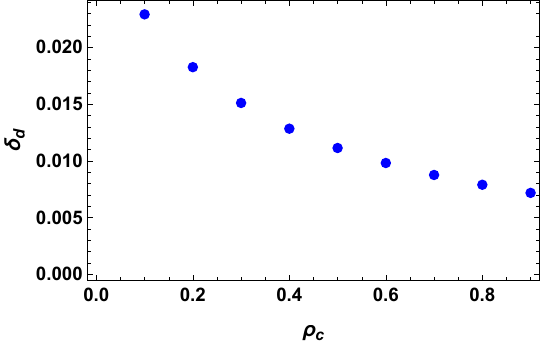}}
\subfigure[\tiny][$\rho_{c}=a=0.5$]{\label{b2}\includegraphics[width=5cm,height=5cm]{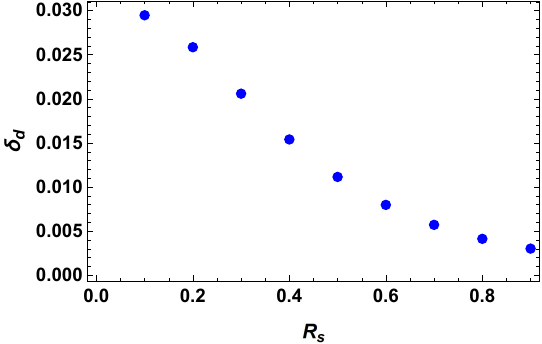}}
\caption{The observable $R_d$ and $\delta_d$ for $r_{obs}=100$,
$\theta_{obs}=90^{\circ}$ and $M=1$.}\label{s1sf3}
\end{center}
\end{figure}
Closely followed by \cite{sd26,sd35}, we characterize the shadow of
rotating BH with the help of two observable measurements, such as the
radius $R_{d}$ and the distortion parameter $\delta_{d}$. For better
understanding, the readers can see Ref. \cite{sd35},
where the authors explained this setup through the schematic
diagram. In Fig. \textbf{\ref{s1sf3}}, we illustrate the influence
of the spin parameter $a$, critical density $\rho_{c}$ and scale
radius $R_{s}$, on $R_{d}$ and $\delta_{d}$. The results indicate
that the radius $R_{d}$ increases with the increasing of
$a$,~$\rho_{c}$ and $R_{s}$, while the distortion $\delta_{d}$
decreases with the increasing of both $\rho_{c}$ and $R_{s}$.
However, $\delta_{d}$ increases with respect to the variation of
$a$. These results conclude that the shadow radius becomes larger
and closer to the standard precise circular shape of the shadow. In
this way, one can observe the boundary of rotating BH shadow on the
observer's screen. We have considered different values for the
parameters $a$,~$\rho_{c}$ and $R_{s}$ for obtaining the rotating BH
shadow contours that have been depicted in Fig. \textbf{\ref{sf2}}.
From the left panel of Fig. \textbf{\ref{sf2}}, we have observed
that the BH shadow contours are shifted rightwards with the
increasing values of spin parameter $a$. Moreover, the widths
between the circular orbits are more obvious on the left side of the
screen as compared to the right side, and when $a=1.5$, the circular shape
of shadow contours is slightly deformed into a D-shape. The influence
of critical density $\rho_{c}$ and the scale radius of the DM halo
$R_{s}$ are interpreted in the middle and right panels of
Fig. \textbf{\ref{sf2}}. From these panels, one can see that the
increasing values of both parameters $\rho_{c}$ and $R_{s}$
increase the radius of circular orbits of the BH shadow. Further, both
parameters do not influence the deformation of the shadow.

\section{Optical Images in the Background of Celestial Light Source}

\begin{figure}[H]
\begin{center}
\subfigure[\tiny][$R_{s}=0.1,~\rho_{c}=0.05$]{\label{a1}\includegraphics[width=3.9cm,height=4cm]{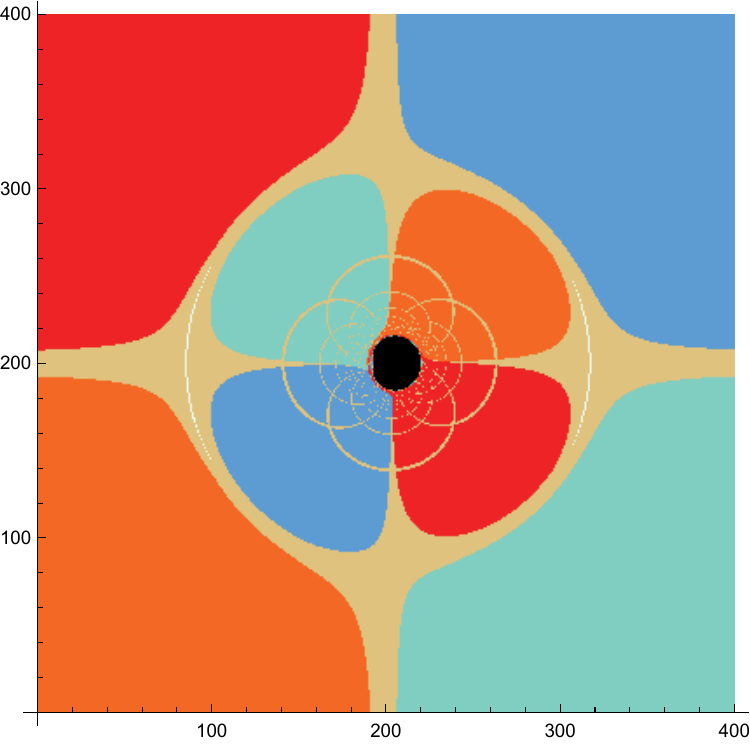}}
\subfigure[\tiny][$R_{s}=0.1,~\rho_{c}=0.25$]{\label{b1}\includegraphics[width=3.9cm,height=4cm]{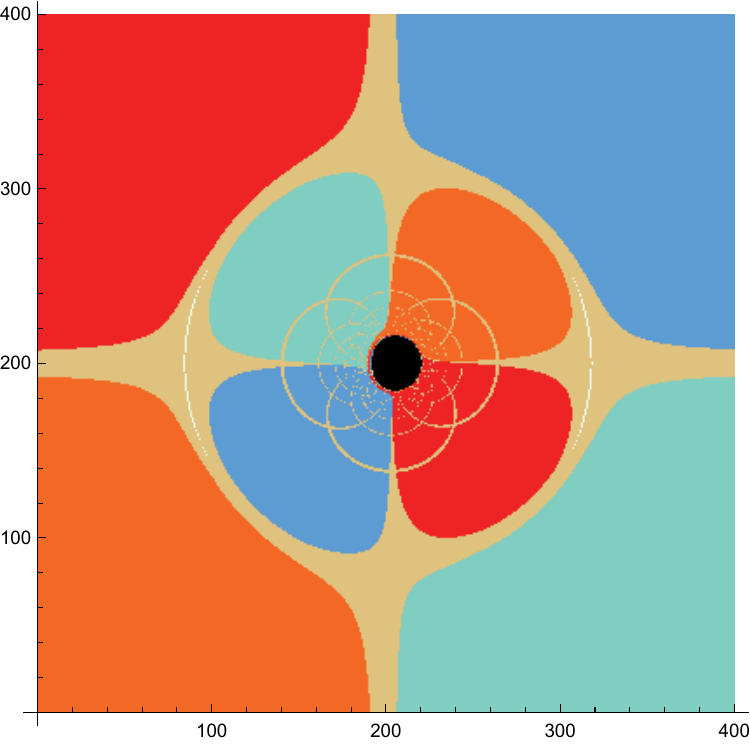}}
\subfigure[\tiny][$R_{s}=0.1,~\rho_{c}=0.45$]{\label{c1}\includegraphics[width=3.9cm,height=4cm]{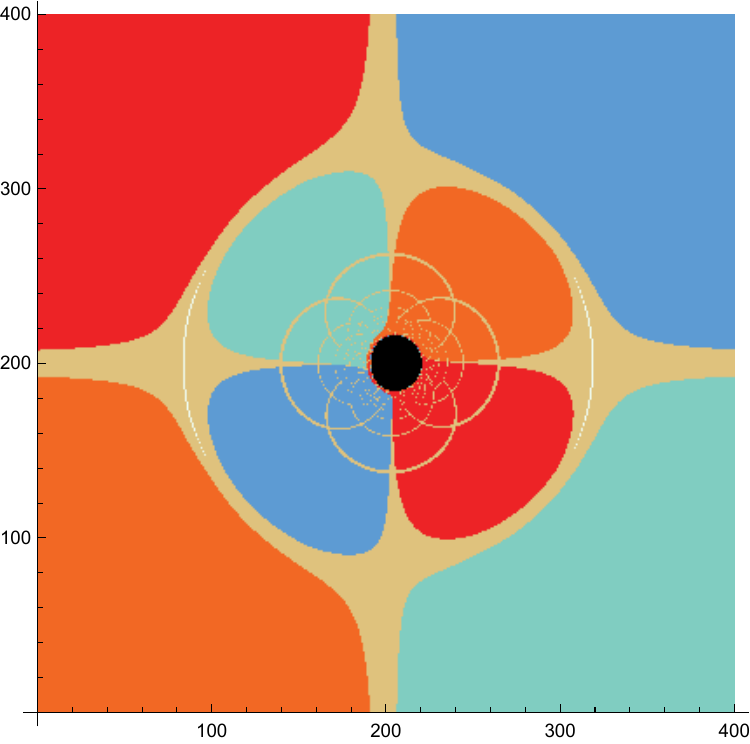}}
\subfigure[\tiny][$R_{s}=0.1,~\rho_{c}=0.65$]{\label{d1}\includegraphics[width=3.9cm,height=4cm]{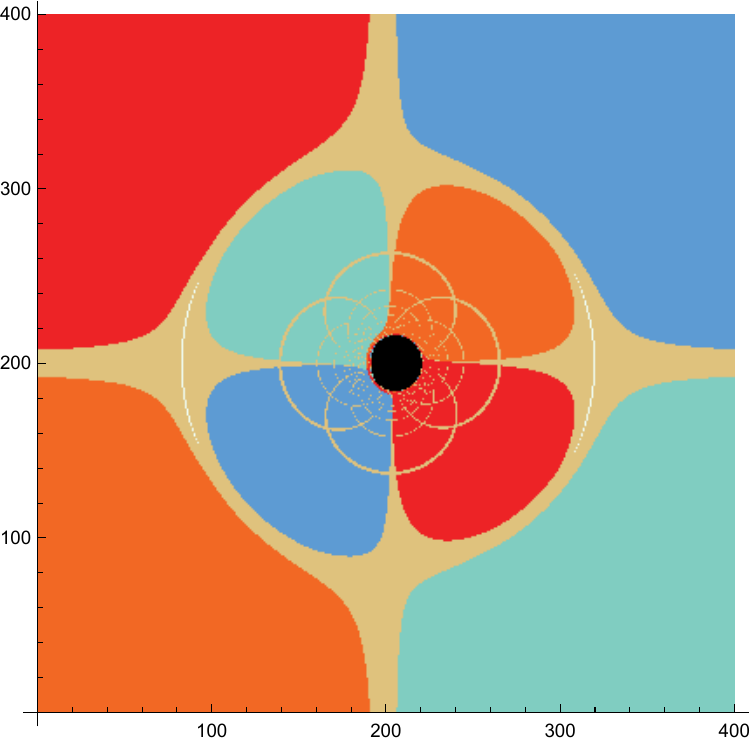}}
\subfigure[\tiny][$R_{s}=0.2,~\rho_{c}=0.05$]{\label{a2}\includegraphics[width=3.9cm,height=4cm]{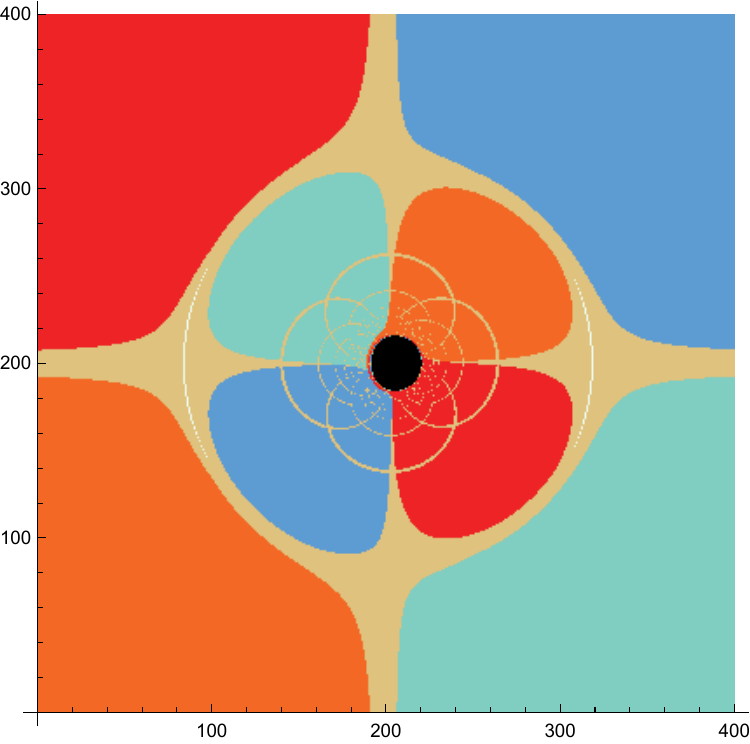}}
\subfigure[\tiny][$R_{s}=0.2,~\rho_{c}=0.25$]{\label{b2}\includegraphics[width=3.9cm,height=4cm]{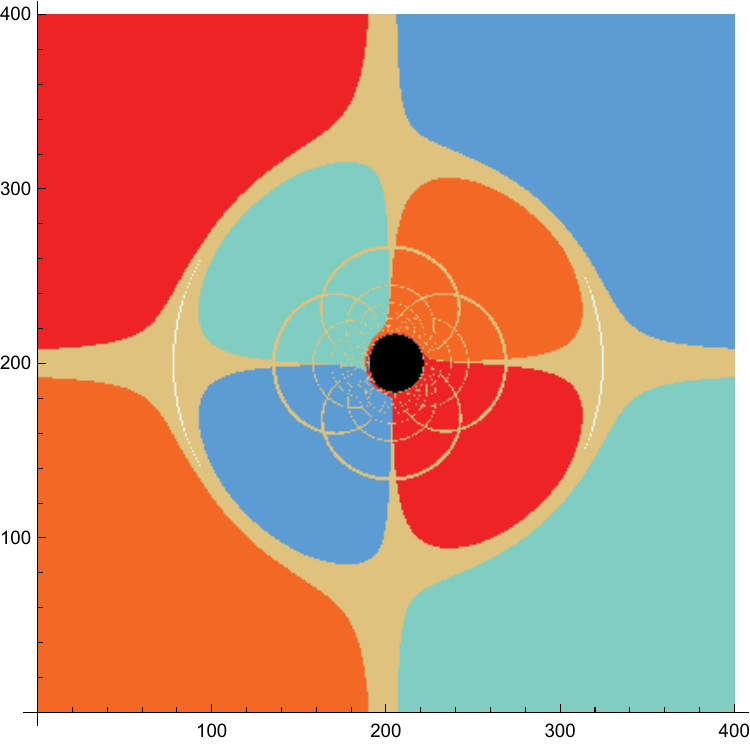}}
\subfigure[\tiny][$R_{s}=0.2,~\rho_{c}=0.45$]{\label{d2}\includegraphics[width=3.9cm,height=4cm]{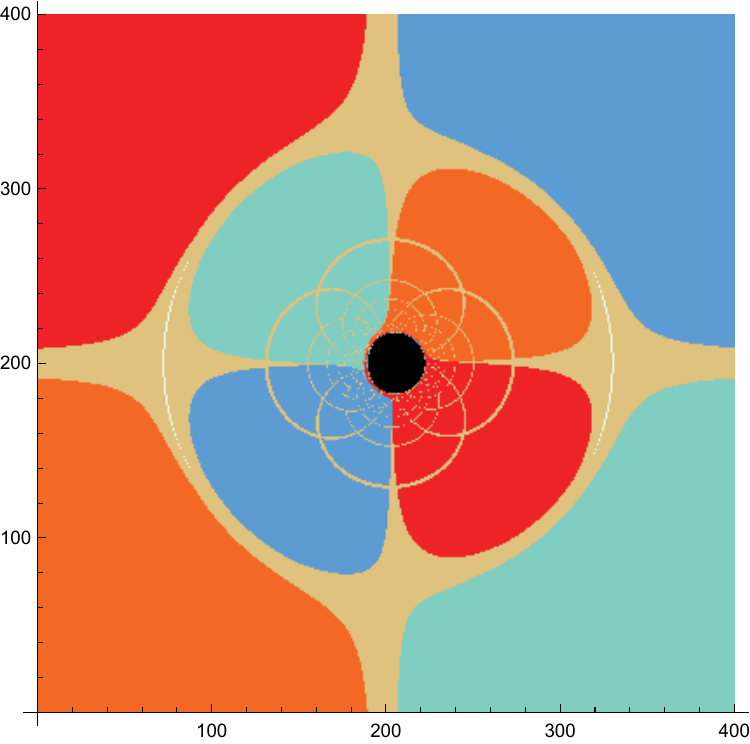}}
\subfigure[\tiny][$R_{s}=0.2,~\rho_{c}=0.65$]{\label{d2}\includegraphics[width=3.9cm,height=4cm]{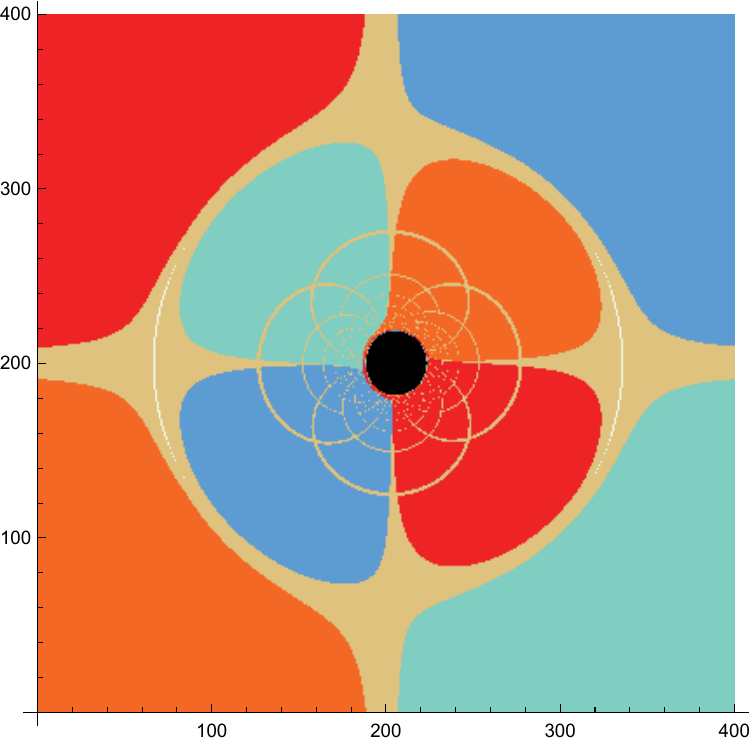}}
\subfigure[\tiny][$R_{s}=0.3,~\rho_{c}=0.05$]{\label{a3}\includegraphics[width=3.9cm,height=4cm]{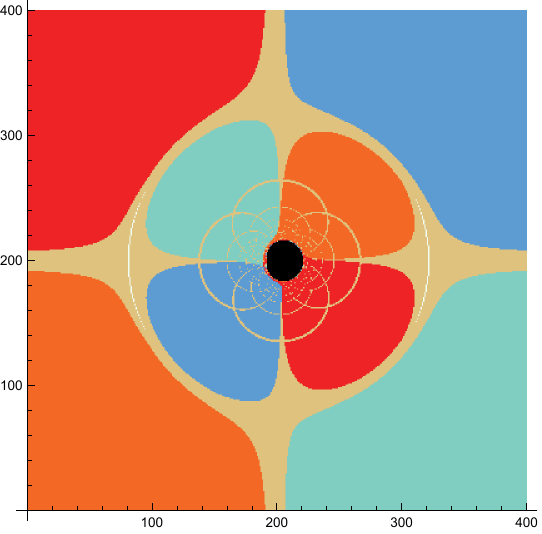}}
\subfigure[\tiny][$R_{s}=0.3,~\rho_{c}=0.25$]{\label{b3}\includegraphics[width=3.9cm,height=4cm]{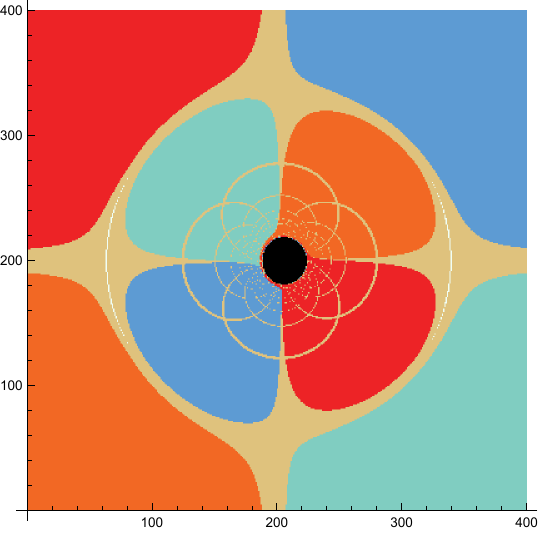}}
\subfigure[\tiny][$R_{s}=0.3,~\rho_{c}=0.45$]{\label{c3}\includegraphics[width=3.9cm,height=4cm]{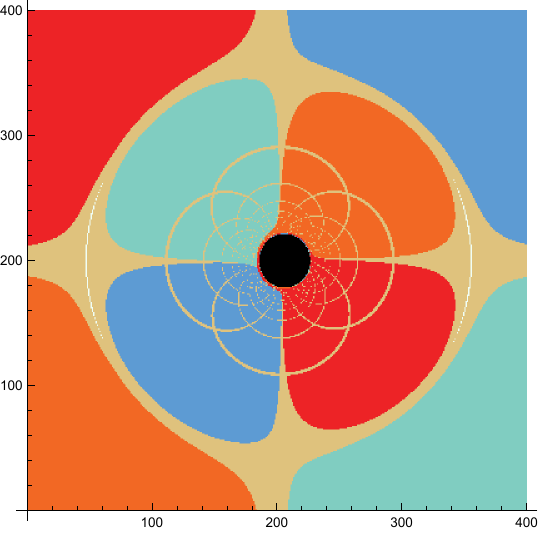}}
\subfigure[\tiny][$R_{s}=0.3,~\rho_{c}=0.65$]{\label{d3}\includegraphics[width=3.9cm,height=4cm]{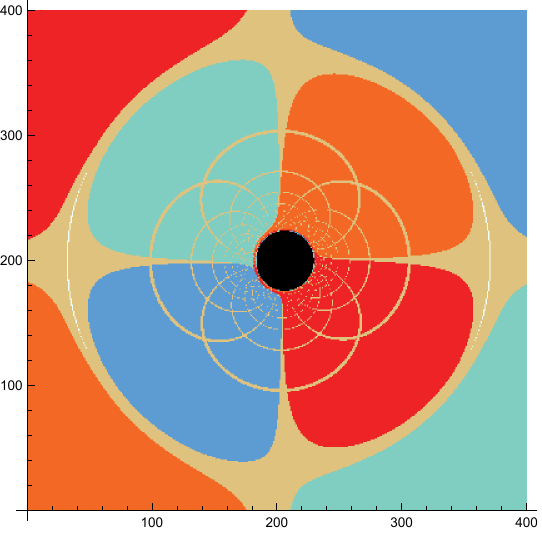}}
\caption{Shadows cast by rotating BH in the presence of CDM halo
under several values of $\rho_{c}$ and $R_{s}$ with $a=0.998$ and
$\theta_{O}=60^{\circ}$. Further, the horizontal and vertical axis
correspond to $x/M$ and $y/M$, respectively.}\label{sf3}
\end{center}
\end{figure}

Now, we impose the backward ray-tracing procedure to explore the
image of BH shadow within the realm of a celestial light source. In
this model, the solid black disk, which represents the BH shadow, is
located in the centre of the celestial sphere, with its size being
remarkably smaller than both the sphere and the distance between the
observer and the origin. For better physical interpretation, we
divide the celestial sphere into four different parts, which
correspond to four different colours (cyan, orange, red, blue)
having the particular angular ranges. Moreover, the backward
ray-tracing procedure involves tracing lesser light rays, excluding
those that are invisible to the observer. It provides a feasible way
to determine the BH shadow images. Now, closely followed by the
strategy as defined in \cite{sd62}, we impose the fish eye camera
model and obtain the BH shadow images for different values of
$\rho_{c}$ and $R_{s}$, as shown in Fig. \textbf{\ref{sf3}} with
fixed values of $a=0.998$ and $\theta_{obs}=60^{\circ}$. All these
images consistently interpret the dark area in the centre, and there
will be an arc-like shape outside the shadow, which is so-called
Einstein ring. Obviously, these images show the warping of space by
a BH and the gravitational lensing phenomenon of a BH.

In Fig. \textbf{\ref{sf3}} (first row), the correlation effect of a
change in the critical density $\rho_{c}$ on the BH shadow is
depicted, where the parameters $\rho_{c}=0.05,~0.25,~0.45,~0.65$ and
$R_{s}=0.1$ is considered. The obtained results showed that with the
increasing values of $\rho_{c}$, there is no significant effect on
deformation of the shadow, and D-shape colours maintain the
optical appearance indicating that there is no drag effect in this
space-time. However, with the augmentation of $\rho_{c}$, the radius
of Einstein ring is slightly increased. In the second and third
rows of Fig. \textbf{\ref{sf3}}, we increase the values of
$R_{s}=0.2$ and $R_{s}=0.3$, respectively, where the values of
$\rho_{c}$ remains the same as defined in the first row of Fig.
\textbf{\ref{sf3}}. In both cases, we notice that with the aid of
$\rho_{c}$, the shape of the inner shadow region slightly evolves
into a D-shape, which aligns with the previous case. Consequently,
white arcs appear outside the D-shape petals on the left and
right side of the screen, which corresponds to the resulting
Einstein ring. Moreover, when both $R_{s}$ and $\rho_{c}$ has
smaller values, the shadow shape of the BH manifests as a D-shape,
however, when both values are increased, the shape of BH gradually
deviates from a D-shape and finally interprets a precise circle.
Moreover, there is a significant increase in the size of the shadow
as well as the vertical axis, when $R_{s}=0.3$ and $\rho_{c}=0.65$.

\section{Thin Accretion Disk Model and Imaging}

\begin{figure}[H]
\centering
\includegraphics[width=18cm,height=7.5cm]{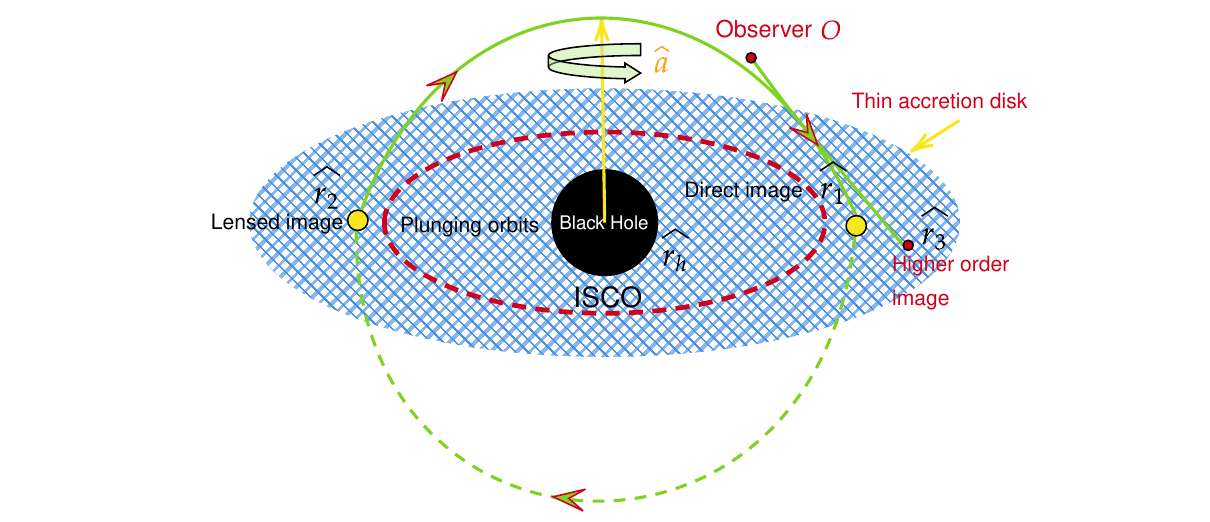}
\caption{The schematic illustration of a BH imaging, where the solid
black disk indicates the BH, the red dashed line represents the ISCO
and the blue elliptical disk denotes the thin accretion disk.
Further, the complete path of light which is received by the distant
observer indicated by the blue curve.}\label{sf4}
\end{figure}

In a real universe scenario, BHs are surrounded by large amounts of
accreting matter, which is accelerated owing to the gravitational force
of the BH. This energetic movement produces high energy radiation
due to friction and heating. As a consequence, a luminous accretion
disk exists close to the BH. In this view, we use the accretion disk
model as outlined in \cite{sd29} to further analyse the observable
features of a rotating BH in CDM halo with a thin accretion disk. For
convenience, we consider that the accretion disk is geometrically
and optically thin, lies on the equatorial plane, and the observer
position lies far away from the BH. Further, the flow of accretion
disk can split into two portions such as the innermost stable
circular orbit (ISCO), where the particles lie inside the ISCO, the
accretion disk experiences plunging motion, and the region beyond the
ISCO, where stable circular orbits exist. Moreover, as defined
in \cite{sd29}, the light rays crossing the equatorial plane once,
twice, three or even many times, will give rise to different
appearances of the BH to a distant observer. In this scenario, we
represent the first, second and third intersections of light rays
with $\hat{r_{1}}$,~$\hat{r_{2}}$ and $\hat{r_{3}}$, respectively.
In Fig. \textbf{\ref{sf4}}, we illustrate the schematic diagram of
this setup, where the positions of $\hat{r_{1}}$,~$\hat{r_{2}}$ and
$\hat{r_{3}}$ correspond to direct, lensed and higher order images,
respectively.
In the present study, we consider that the accretion disk begins
from the BH event horizon $\hat{r}_{h}$ and expands to a reasonable
distant point $\hat{r}_{f}$, which we set $\hat{r}_{f}=1000$. And
the observer's position lies in the region
$\hat{r}_{h}\ll\hat{r}_{obs}<\hat{r}_{f}$. Now for the desired
results, we first need to find the radius of the ISCO, which can be
determined with the help of the following equations
\cite{sd29,sd30,sd63}
\begin{eqnarray}\label{s11}
\mathcal{V}_{e}(r)=0, \quad \partial_{r} \mathcal{V}_{e}(r)=0, \quad
\partial_{r}^{2} \mathcal{V}_{e}(r)=0,
\end{eqnarray}
in which $\mathcal{V}_{e}$ represents the effective potential, which
reads as
\begin{equation}\label{s12}
\mathcal{V}_{e}=(1+g^{tt}\bar{E}^{2}+g^{tt}\bar{L}^{2}-2g^{t\phi}\bar{E}\bar{L}).
\end{equation}
Here $\bar{E}$ and $\bar{L}$ are two constants corresponding to the
particular energy and angular momentum of a massive neutral
particle, respectively. And we have
\begin{eqnarray}\label{s13}
\bar{E}=-\frac{1}{\sqrt{\ell_{1}}}(g_{tt}+g_{t\phi}\widetilde{\Psi}),\quad
\bar{L}=\frac{1}{\sqrt{\ell_{1}}}(g_{t\phi}+g_{\phi\phi}\widetilde{\Psi}),
\end{eqnarray}
where
$\ell_{1}=-g_{tt}-2g_{t\phi}\widetilde{\Psi}-g_{\phi\phi}\widetilde{\Psi}^{2}$
and
$\widetilde{\Psi}=\frac{d\phi}{dt}=(\partial_{r}g_{t\phi}+(\sqrt{\partial^{2}_{r}g_{t\phi}-\partial_{r}g_{tt}\partial_{r}g_{\phi\phi}})(\partial_{r}g_{\phi\phi})^{-1}.$
Now the quantities $\bar{E}$ and $\bar{L}$ read as $\bar{E}_{ISCO}$
and $\bar{L}_{ISCO}$ at ISCO. Outside the ISCO, the accretion flows
move along time-like circular orbits
\begin{equation}\label{s14}
u^{\xi}_{out}=\frac{1}{\sqrt{\ell_{1}}}(1,0,0,\widetilde{\Psi}).
\end{equation}
After that, within the ISCO, the accretion flows descend from the
ISCO to the event on a critical plunging orbits, preserving the
conserved quantities related to the ISCO. In this scenario, the
components of four-velocity are defined as \cite{sd29,sd30,sd63}
\begin{eqnarray}\nonumber
u^{t}_{pl}&=&(-g^{tt}\bar{E}_{ISCO}+g^{t\phi}\bar{L}_{ISCO}),\quad
u^{\phi}_{pl}=(-g^{t\phi}\bar{E}_{ISCO}+g^{\phi\phi}\bar{L}_{ISCO}),\\\nonumber
u^{r}_{pl}&=&-\big(-(g_{tt}u^{t}_{pl}u^{t}_{pl}+2g_{t\phi}u^{t}_{pl}u^{\phi}_{pl}+g_{\phi\phi}u^{\phi}_{pl}u^{\phi}_{pl}+1)(g_{rr})^{-1}\big)^{\frac{1}{2}},\\\label{s15}~u^{\theta}_{pl}&=&0.
\end{eqnarray}
As we defined earlier, when we traced the light rays backward from
the distant observer, these rays may cross the accretion disk
plane once ($n=1$), twice ($n=2$) or even many times ($n>2$). Every
cross-section grants the observer to receive some extra brightness at
the location of intersection indicated as $r_{n} (x,y)$, which is
well-known as the transfer function. Indeed, the function $r_{n}
(x,y)$ produced the $n^{th}$ image of the accretion disk such that
the value $n=1$ corresponds to the direct image, while the value
$n=2$ or $n=3$ corresponds to the lensed image or higher order
images, respectively. Thus, ignoring the influence of reflection and
thickness, the observed intensity of the light enhances every time,
when a light ray crossed the accretion disk. Hence, the
observed intensity on the observer's screen can be defined as
\cite{sd29,sd30,sd63}
\begin{equation}\label{s16}
I_{\nu_{obs}}=\sum_{n=1}^{N_{max}}f_{n}g_{n}^{3}(r_{n})J_{n},
\end{equation}
where $\nu_{obs}$ is the observed frequency on the screen, $N_{max}$
represents the maximum number of times that light rays crossed the
accretion disk, $g_{n}$ is the red-shift factor, $J_{n}$ is the
emissivity factor at the $n^{th}$ intersection and $f_{n}=1$ is a
fudge factor. The expression of $J_{n}$ is defined as
\begin{equation}\label{s17}
J_{n}=\exp\big[\tau_{1}k^{2}+\tau_{2}k\big],
\end{equation}
where $k=\log(\frac{r}{\hat{r_{h}}})$ and $\tau_{1}=-1/2$ and
$\tau_{2}=-2$, which is consistent with $230$ GHz images as captured
by the EHT \cite{sd64}. Further, the red-shift factor
$g_{n}=\nu_{O}/\nu_{n}$, in which $\nu_{n}$ indicates the frequency
observed by the local static frames comoving with the energetic
emission. Naturally, because of the notable differences in particle
emission spectra between both the inner and outer zones of the ISCO,
specific forms of the red-shift factor vary in both of these
domains. The red-shift factor beyond the ISCO can be expressed as
\cite{sd39}
\begin{equation}\label{s18}
g^{out}_{n}=\frac{\varpi(1-\lambda\frac{p_{\phi}}{p_{t}})}{\omega(1+\widetilde{\Psi}\frac{p_{\phi}}{p_{t}})}|_{r=r_{n}},
\quad \quad r\geq r_{ISCO},
\end{equation}
where
$\varpi=\sqrt{\frac{g_{\phi\phi}}{g^2_{t\phi}-g_{tt}g_{\phi\phi}}}$,~$\lambda=\frac{g_{t\phi}}{g_{\phi\phi}}$,~
$\omega=\frac{1}{\sqrt{\ell_{1}}}$ and
$\bar{e}=\frac{p_{(t)}}{p_{t}}=\varpi(1-\lambda\frac{p_{\phi}}{p_{t}})$
is the relationship between the observed energy on the screen to the
energy along a null geodesic. Under these circumstances, we assume
that $\bar{e}=1$ \cite{sd29}. While inside the ISCO, the accretion
flow is moving along the critical plunging orbit, then the red-shift
factor is defined as \cite{sd39}
\begin{equation}\label{s19}
g^{pl}_{n}=-\frac{1}{u^{r}_{pl}p_{r}/p_{t}-\bar{E}_{ISCO}(g^{tt}-g^{t\phi}p_{\phi}/p_{t})
+\bar{L}_{ISCO}(g^{\phi\phi}p_{\phi}/p_{t}+g^{t\phi})}|_{r=r_{n}},
\quad \quad r< r_{ISCO}.
\end{equation}

\subsection{Visual Representation of Rotating Black Hole Enveloped by Cold Dark Matter}

\begin{figure}[H]
\begin{center}
\subfigure[\tiny][$R_{s}=0.1,~\rho_{c}=0.05$]{\label{a1}\includegraphics[width=3.9cm,height=4cm]{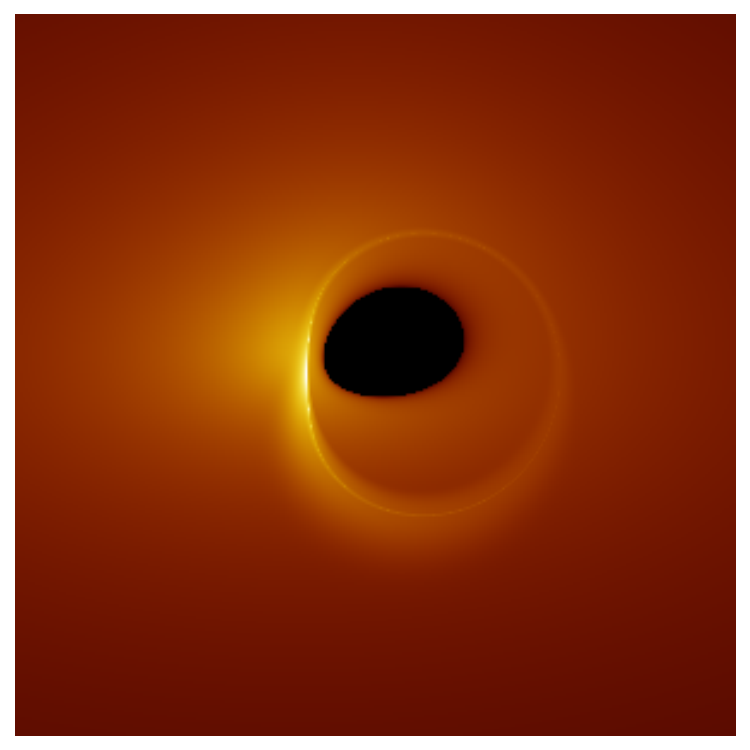}}
\subfigure[\tiny][$R_{s}=0.1,~\rho_{c}=0.25$]{\label{b1}\includegraphics[width=3.9cm,height=4cm]{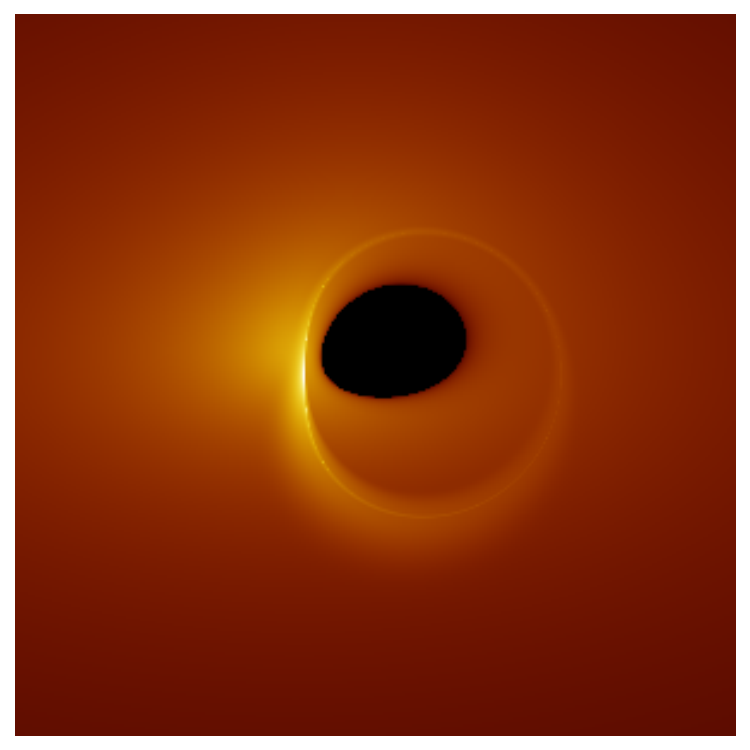}}
\subfigure[\tiny][$R_{s}=0.1,~\rho_{c}=0.45$]{\label{c1}\includegraphics[width=3.9cm,height=4cm]{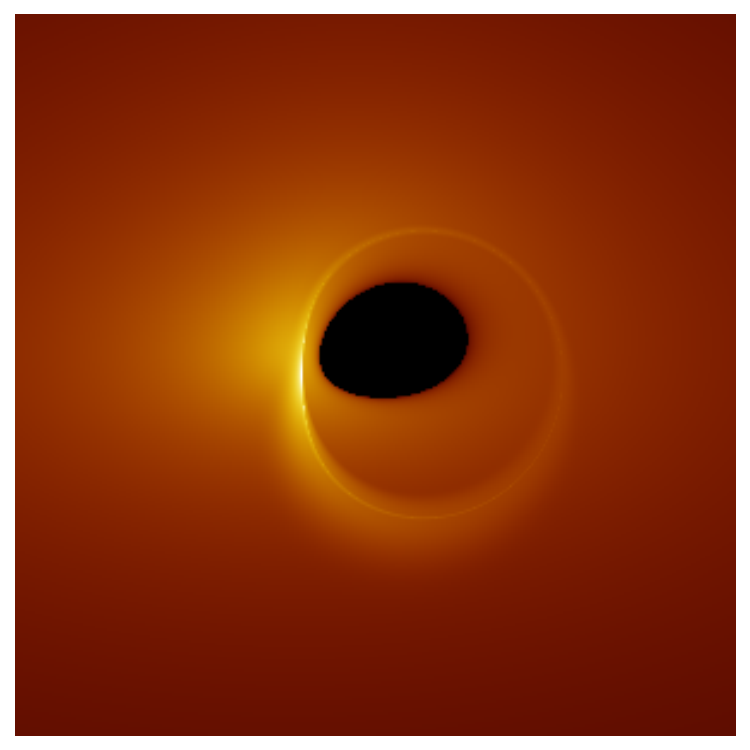}}
\subfigure[\tiny][$R_{s}=0.1,~\rho_{c}=0.65$]{\label{d1}\includegraphics[width=3.9cm,height=4cm]{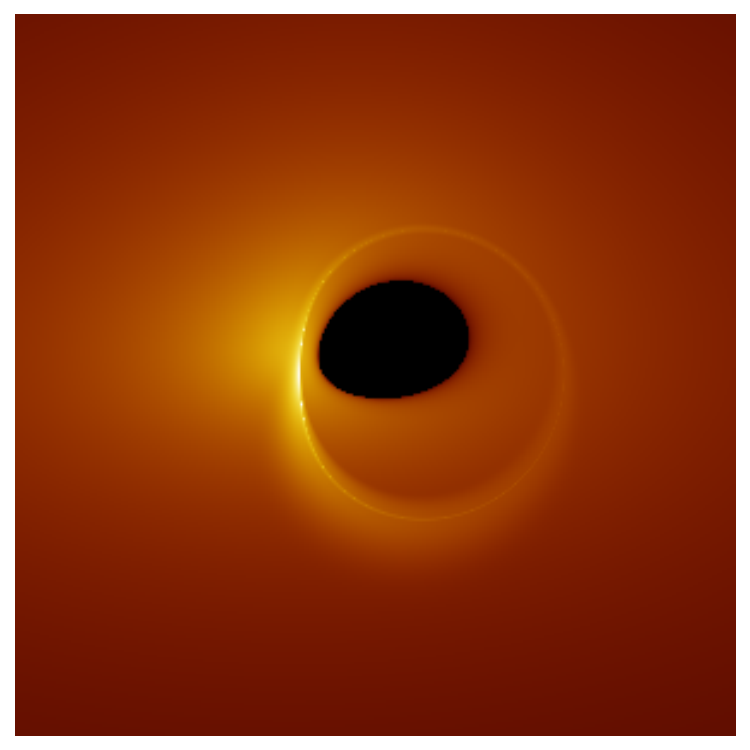}}
\subfigure[\tiny][$R_{s}=0.2,~\rho_{c}=0.05$]{\label{a2}\includegraphics[width=3.9cm,height=4cm]{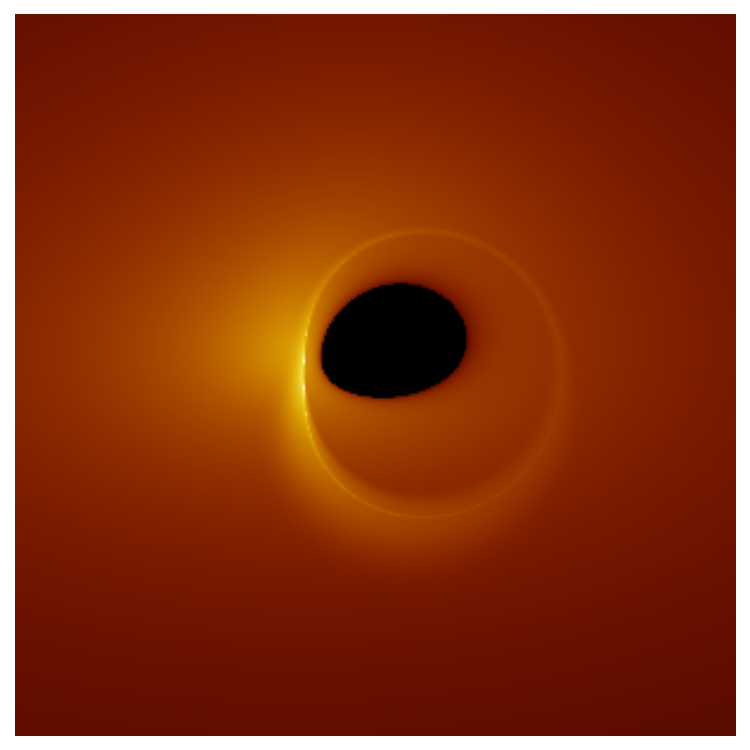}}
\subfigure[\tiny][$R_{s}=0.2,~\rho_{c}=0.25$]{\label{b2}\includegraphics[width=3.9cm,height=4cm]{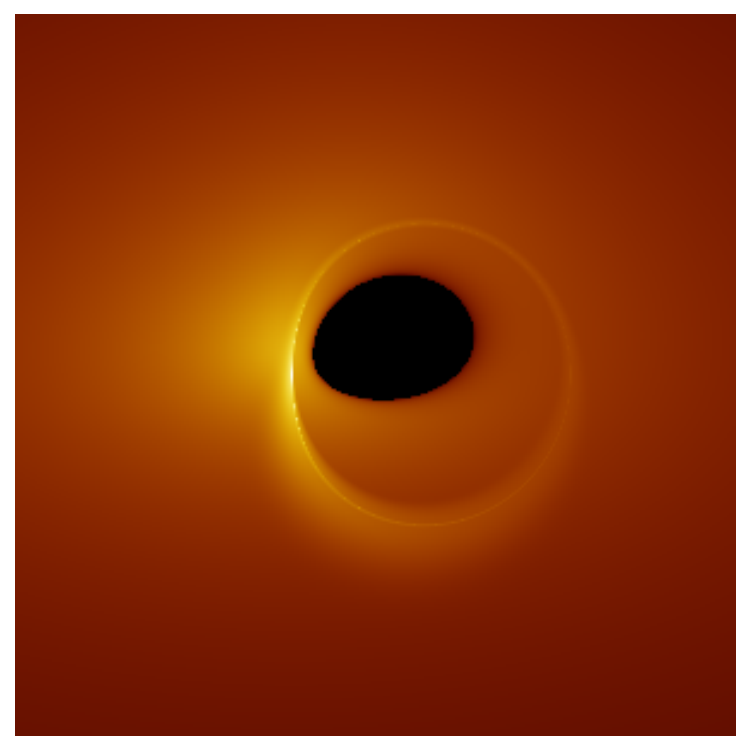}}
\subfigure[\tiny][$R_{s}=0.2,~\rho_{c}=0.45$]{\label{d2}\includegraphics[width=3.9cm,height=4cm]{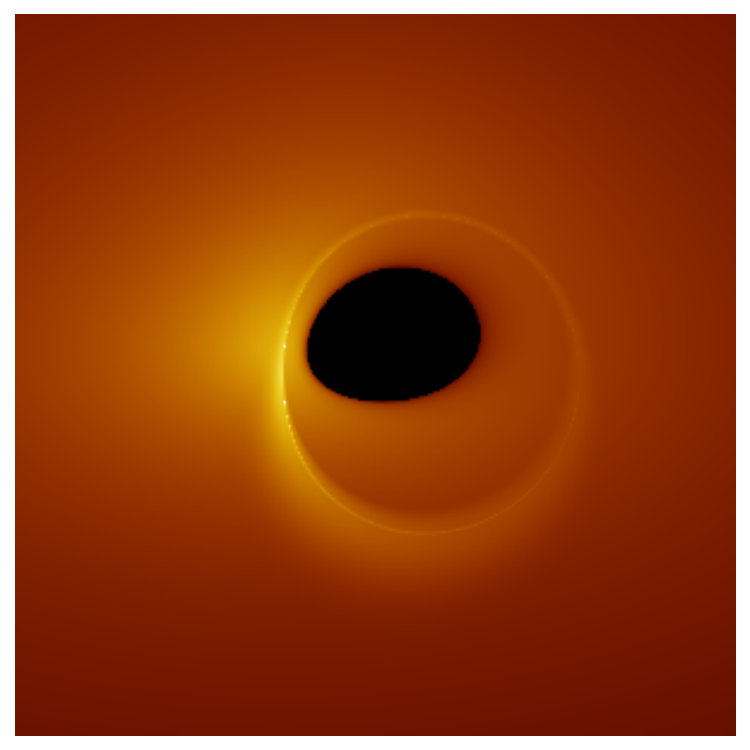}}
\subfigure[\tiny][$R_{s}=0.2,~\rho_{c}=0.65$]{\label{d2}\includegraphics[width=3.9cm,height=4cm]{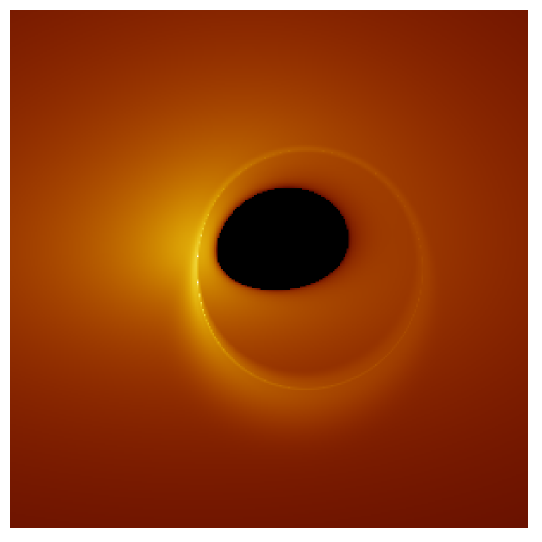}}
\subfigure[\tiny][$R_{s}=0.3,~\rho_{c}=0.05$]{\label{a3}\includegraphics[width=3.9cm,height=4cm]{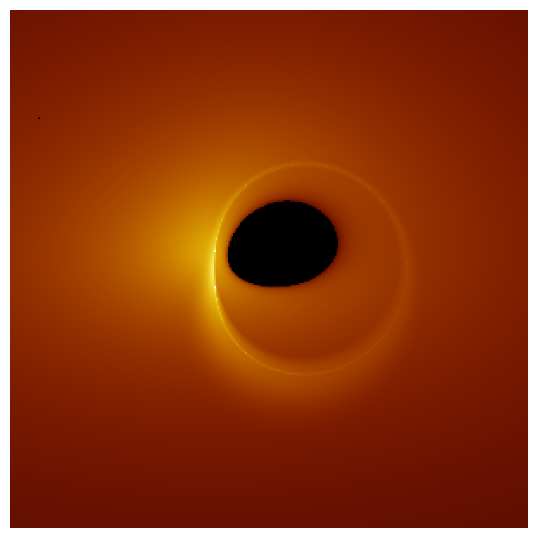}}
\subfigure[\tiny][$R_{s}=0.3,~\rho_{c}=0.25$]{\label{b3}\includegraphics[width=3.9cm,height=4cm]{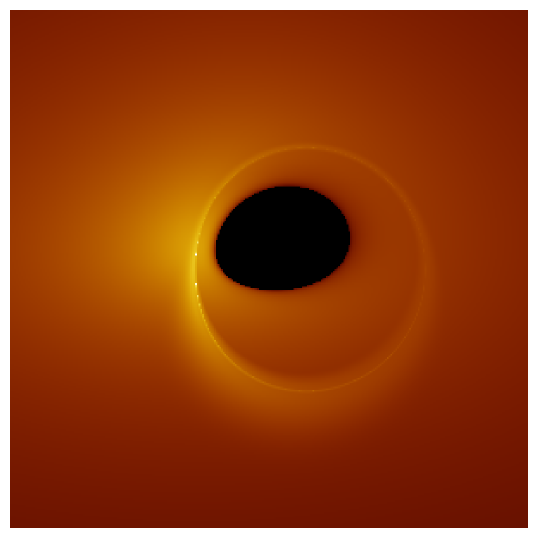}}
\subfigure[\tiny][$R_{s}=0.3,~\rho_{c}=0.45$]{\label{c3}\includegraphics[width=3.9cm,height=4cm]{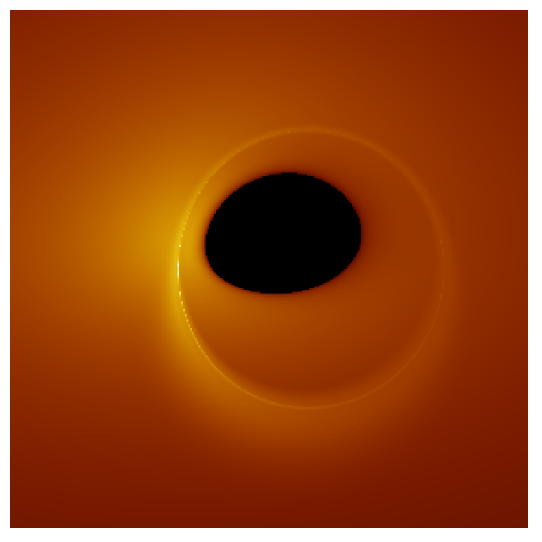}}
\subfigure[\tiny][$R_{s}=0.3,~\rho_{c}=0.65$]{\label{d3}\includegraphics[width=3.9cm,height=4cm]{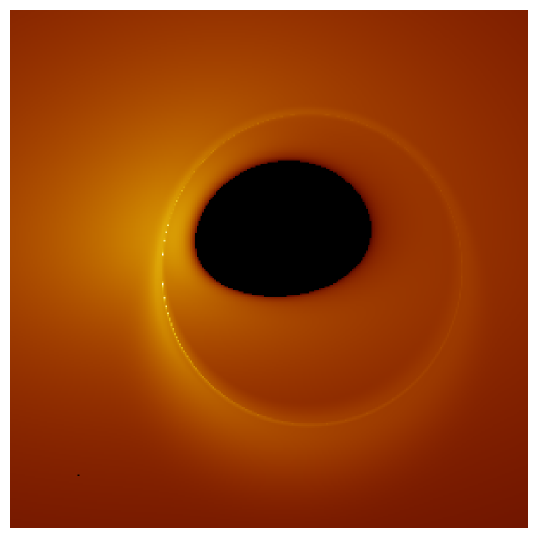}}
\caption{Shadows cast by rotating BH in the presence of CDM halo
surrounded by prograde flow at $230$ GHz under several values of
$\rho_{c}$ and $R_{s}$ with $a=0.998$ and $\theta_{obs}=60^{\circ}$.
Further, the horizontal and vertical axis correspond to $x/M$ and
$y/M$, respectively.}\label{sf5}
\end{center}
\end{figure}

After specifying the attributes of the thin accretion disk model and
measuring the observable intensity obtained by the observer, we may
simulate a representation of the rotating BH within the realm
of CDM halo lighted by the accretion disk on the observer frame. We
investigate two different forms of accretion flow behaviour, such as
prograde (forward photons) accretion flow and retrograde (backward
photons) accretion flow, in contrast to the rotating direction of
considering space-time. In Fig. \textbf{\ref{sf5}}, we interpret
the optical appearance of a rotating BH in CDM halo surrounded by
prograde flows under different values of scale radius $R_{s}$ and
critical density $\rho_{c}$ with fixed values of spin parameter
$a=0.998$ and $\theta_{obs}=60^{\circ}$. In all images, the bright
photon rings are observed as narrow curves, which happens due to
the influence of strong gravitational lensing. The optical
appearance of these intensity images shows that there is a central
dark region, which is the result of the direct images of horizons,
known as the inner shadow \cite{sd64}. In order to illustrate the
influence of critical density $\rho_{c}$, we fixed the values of
scale radius $R_{s}$ and vary the values of $\rho_{c}$ from left
to right as $0.05,~0.25,~0.45$ and $0.65$. Whereas from top to
bottom, the intensity maps demonstrate the influence of the $R_{s}$
on the intricate properties of the shadow images. From these images,
it is observed that with the variation of the $\rho_{c}$, the shadow
images deformed, emerging a smooth, small semi-circular black region.
Particularly, when parameter $R_{s}$ is fixed and $\rho_{c}$ is
increased, it shows that the area of the inner shadow contracts and
luminosity moves towards the upper half-portion of the screen. On
the other hand, when $\rho_{c}$ is fixed and the value of $R_{s}$
increased, the size of the inner shadow gradually increases, which
is stronger than the influence of the critical density $\rho_{c}$.
Moreover, with the increases of both $R_{s}$ and $\rho_{c}$, the
overall intensity of the observed light of the image interprets an
upward trend.
To understand how light from the BH's accretion disk reaches the
observer's frame, it is essential to examine the fluctuations in the
light intensity caused by divergence, absorption, Doppler effect,
and gravitational red-shift. The relative motion between the
accretion disk and the distant observers provide insight into the
complex interactions between BHs and accretion disks, and their
development. The phenomenon of the Doppler effect provides an important
aspect in this direction, whereas the gravitational red-shift plays
a significant role in revealing the influence of gravitational
field in this region. Consequently, it is essential to properly
determine the red-shift factor related to the behaviour of radiated
particles during the imaging process of a BH, since this step is
essential to accurately identify the BH and its accretion disk.

\begin{figure}[H]
\begin{center}
\subfigure[\tiny][$R_{s}=0.1,~\rho_{c}=0.05$]{\label{a1}\includegraphics[width=3.9cm,height=4cm]{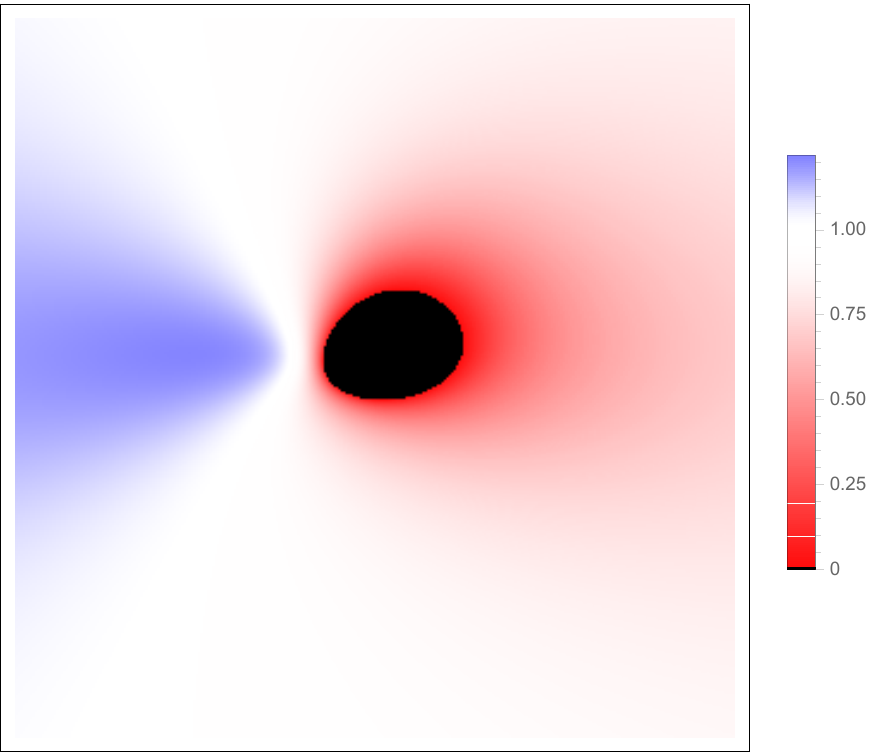}}
\subfigure[\tiny][$R_{s}=0.1,~\rho_{c}=0.25$]{\label{b1}\includegraphics[width=3.9cm,height=4cm]{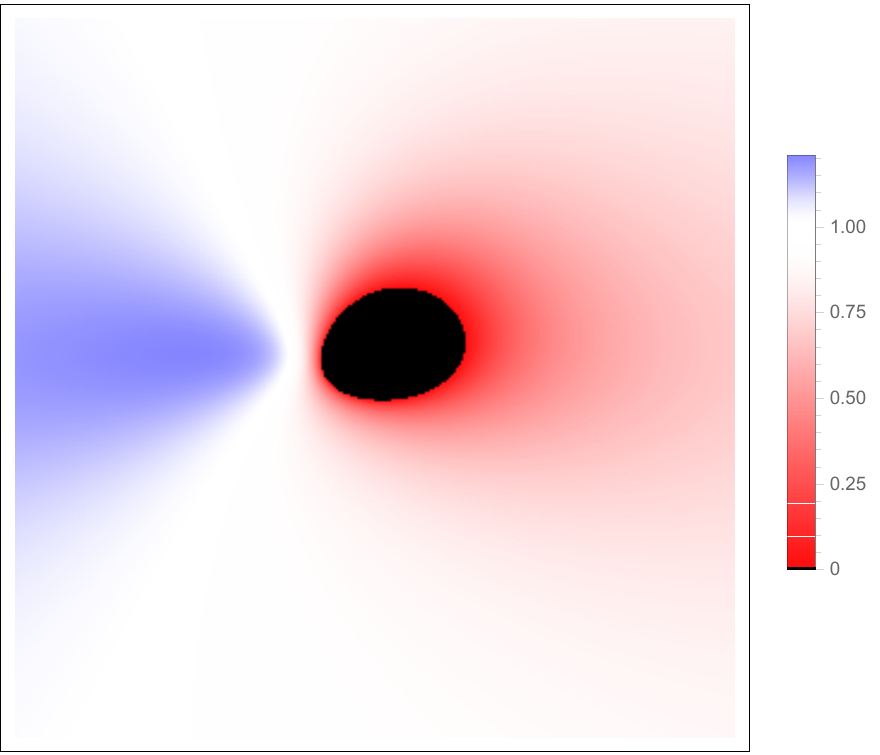}}
\subfigure[\tiny][$R_{s}=0.1,~\rho_{c}=0.45$]{\label{c1}\includegraphics[width=3.9cm,height=4cm]{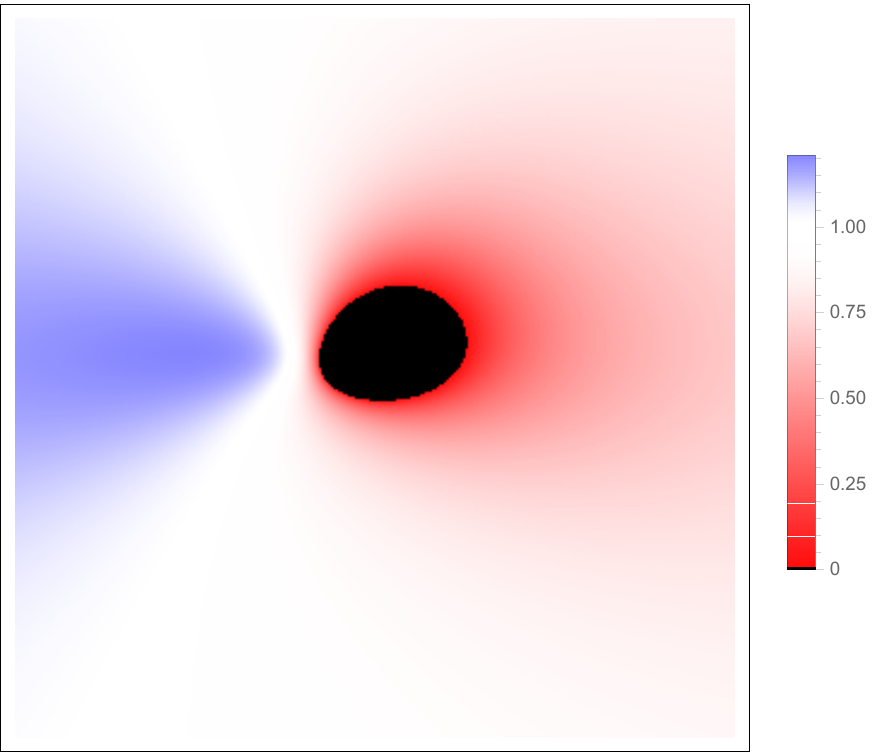}}
\subfigure[\tiny][$R_{s}=0.1,~\rho_{c}=0.65$]{\label{d1}\includegraphics[width=3.9cm,height=4cm]{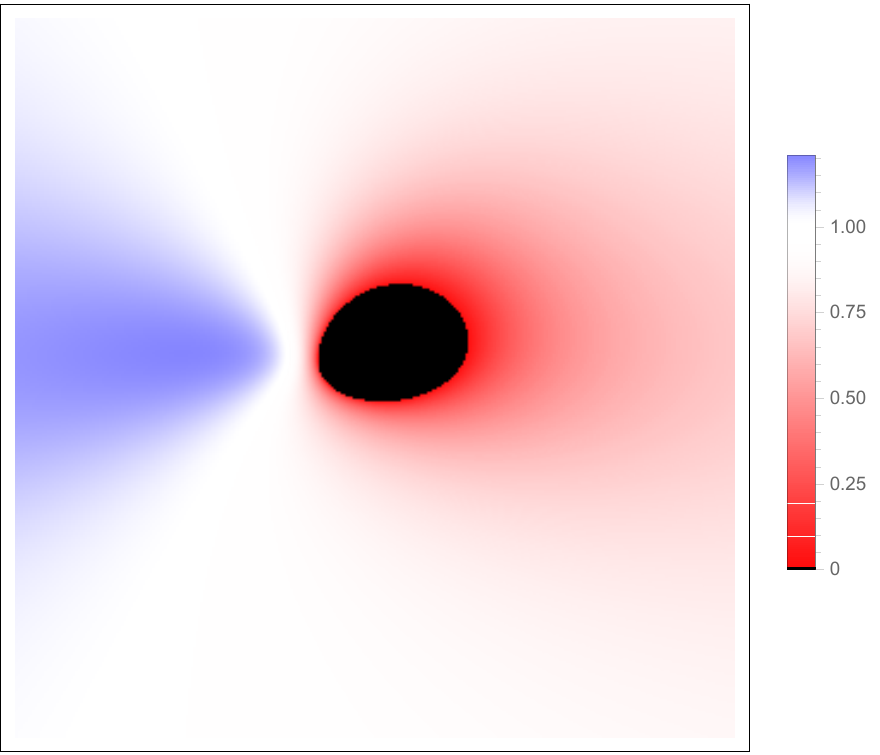}}
\subfigure[\tiny][$R_{s}=0.2,~\rho_{c}=0.05$]{\label{a2}\includegraphics[width=3.9cm,height=4cm]{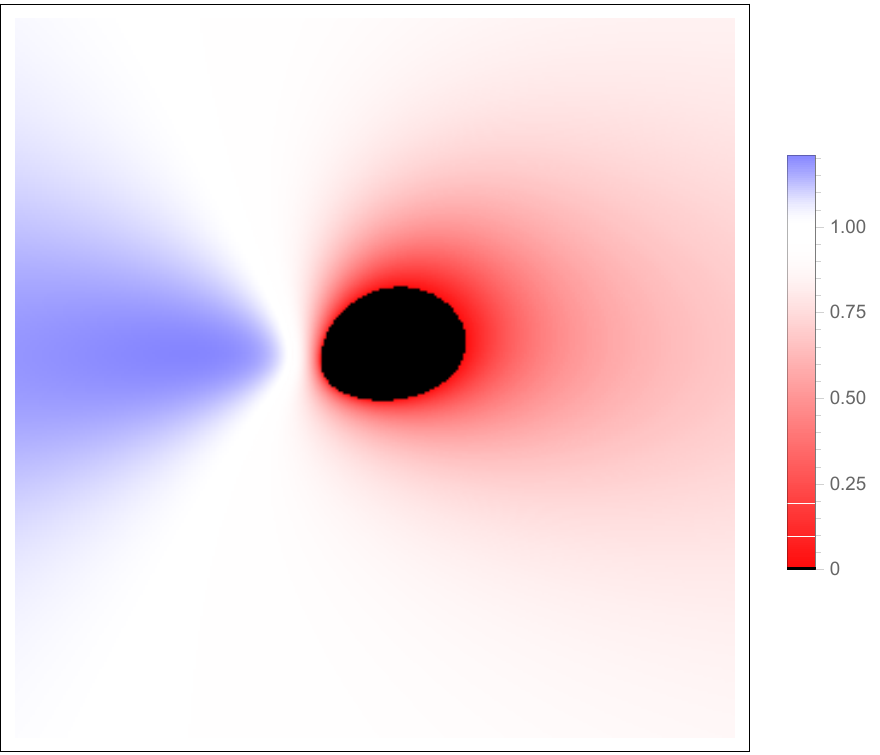}}
\subfigure[\tiny][$R_{s}=0.2,~\rho_{c}=0.25$]{\label{b2}\includegraphics[width=3.9cm,height=4cm]{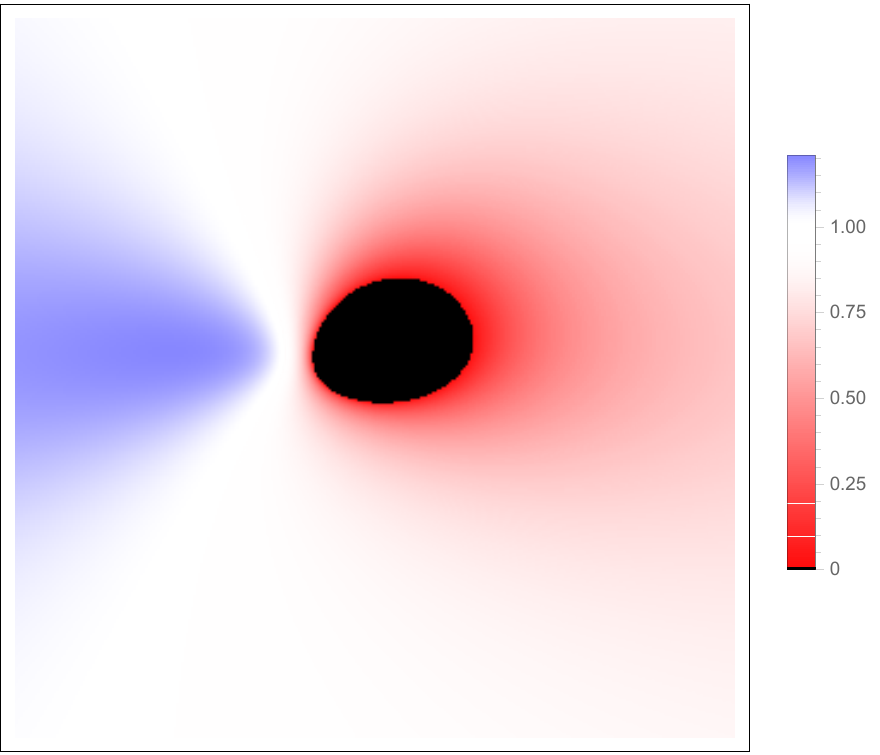}}
\subfigure[\tiny][$R_{s}=0.2,~\rho_{c}=0.45$]{\label{d2}\includegraphics[width=3.9cm,height=4cm]{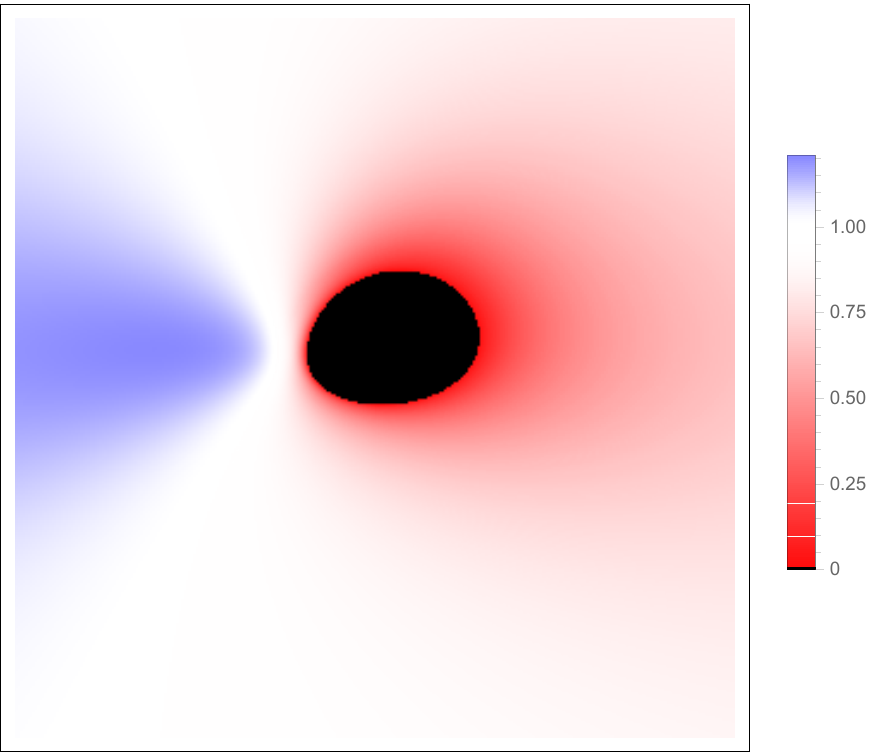}}
\subfigure[\tiny][$R_{s}=0.2,~\rho_{c}=0.65$]{\label{d2}\includegraphics[width=3.9cm,height=4cm]{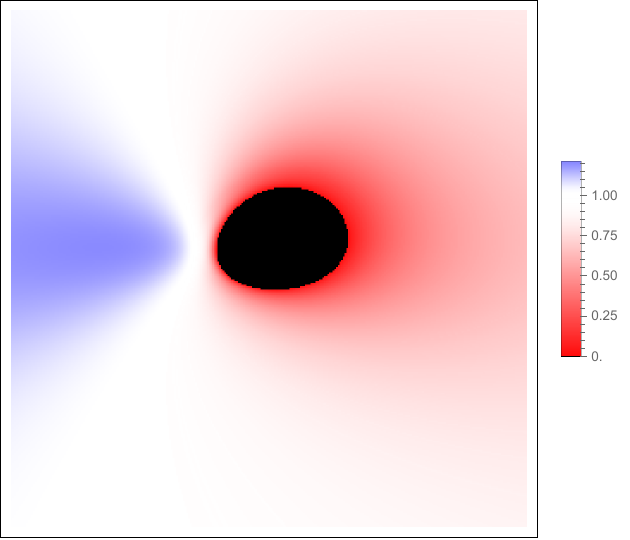}}
\subfigure[\tiny][$R_{s}=0.3,~\rho_{c}=0.05$]{\label{a3}\includegraphics[width=3.9cm,height=4cm]{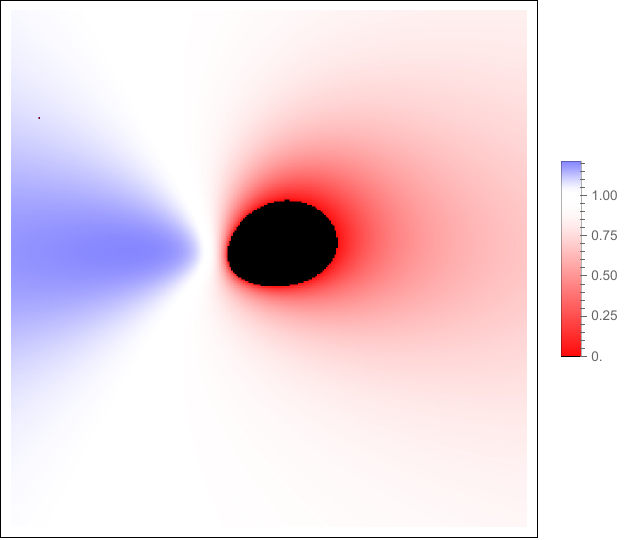}}
\subfigure[\tiny][$R_{s}=0.3,~\rho_{c}=0.25$]{\label{b3}\includegraphics[width=3.9cm,height=4cm]{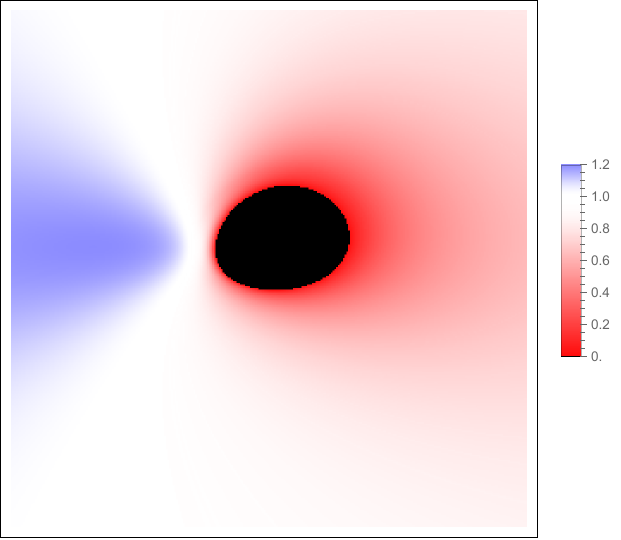}}
\subfigure[\tiny][$R_{s}=0.3,~\rho_{c}=0.45$]{\label{c3}\includegraphics[width=3.9cm,height=4cm]{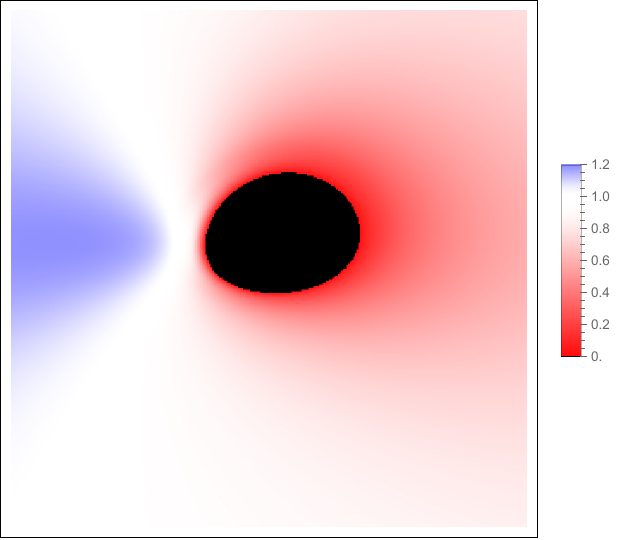}}
\subfigure[\tiny][$R_{s}=0.3,~\rho_{c}=0.65$]{\label{d3}\includegraphics[width=3.9cm,height=4cm]{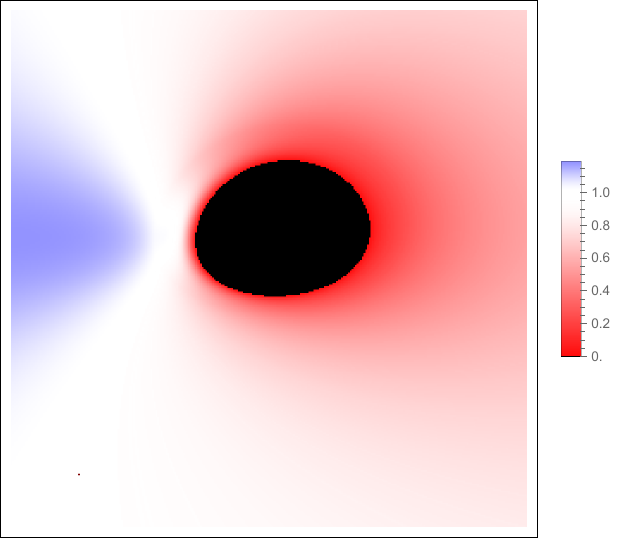}}
\caption{The red-shifts configuration of direct images with prograde
flow under several values of $\rho_{c}$ and $R_{s}$ with $a=0.998$
and $\theta_{obs}=60^{\circ}$. Further, the horizontal and vertical
axis correspond to $x/M$ and $y/M$, respectively.}\label{sf6}
\end{center}
\end{figure}

\begin{table}[H]\centering
\begin{tabular}{|c|c|c|c|c|c|c|c|}
\hline \diagbox{$\rho_c$}{$R_s$} & 0.1 & 0.2 & 0.3 & 0.4 & 0.5 & 0.6
& 0.7 \\ \hline 0.05 & 1.21509 & 1.21431 & 1.21245 & 1.20911 &
1.20407 & 1.19797 & 1.19125 \\ \hline 0.15 & 1.21487 & 1.21276 &
1.20767 & 1.19967 & 1.19058 & 1.17491 & 1.13472 \\ \hline 0.25 &
1.21464 & 1.21132 & 1.20359 & 1.19319 & 1.17548 & 1.13901 & 1.43687
\\ \hline 0.35 & 1.21443 & 1.20988 & 1.20005 & 1.18754 & 1.14602 &
1.40016 & 3.07342 \\ \hline 0.45 & 1.21422 & 1.2085 & 1.19715 &
1.1782 & 1.23141 & 2.0458 & 483.49 \\ \hline 0.55 & 1.214 & 1.20715
& 1.19464 & 1.16448 & 1.4245 & 4.89924 & 0.80747 \\ \hline 0.65 &
1.21382 & 1.20597 & 1.19248 & 1.15103 & 1.76166 & 543.117 & 0.633372
\\ \hline
\end{tabular}
\caption{The maximal blue-shift $g_{max} $ of direct images under
different values of $R_s$ and $\rho_c$ with $a=0.998$ and
$\theta_{obs}=60^{\circ}$.}\label{tab1}
\end{table}

\begin{figure}[H]
\begin{center}
\subfigure[\tiny][$R_{s}=0.1,~\rho_{c}=0.05$]{\label{a1}\includegraphics[width=3.9cm,height=3.8cm]{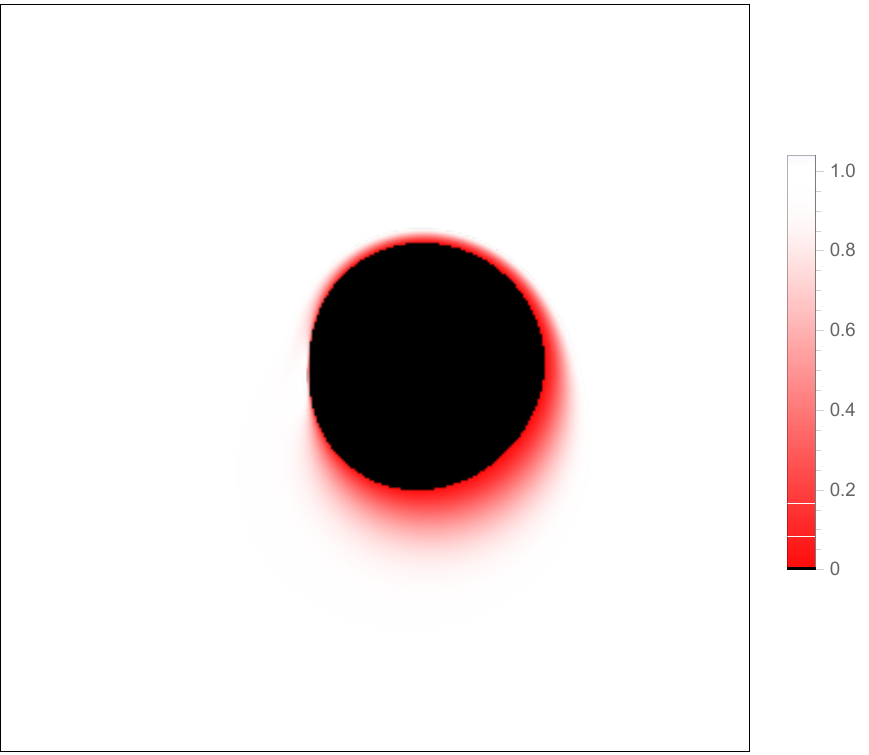}}
\subfigure[\tiny][$R_{s}=0.1,~\rho_{c}=0.25$]{\label{b1}\includegraphics[width=3.9cm,height=3.8cm]{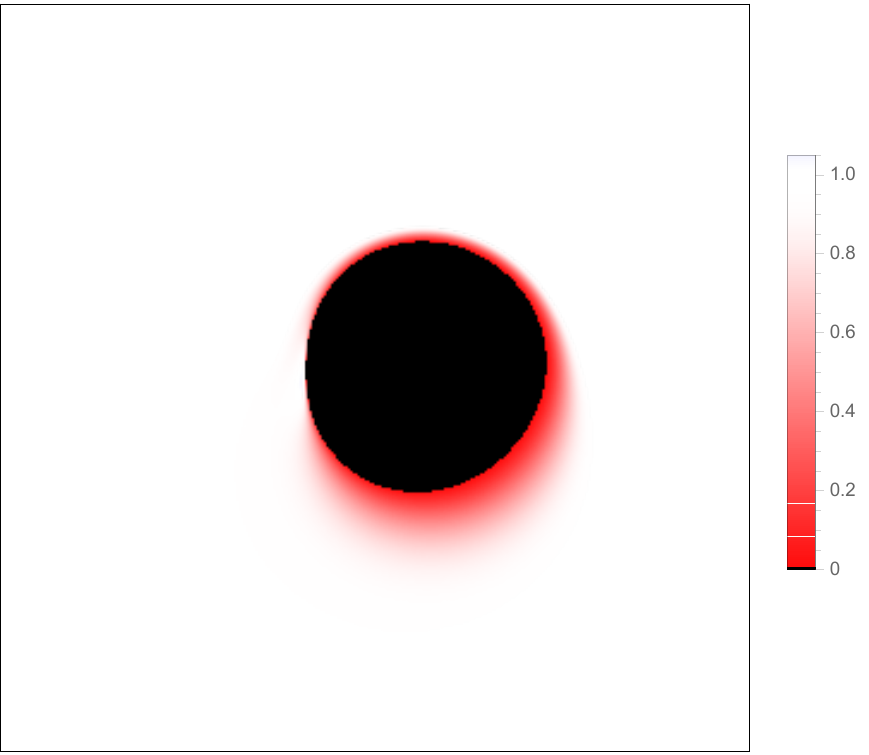}}
\subfigure[\tiny][$R_{s}=0.1,~\rho_{c}=0.45$]{\label{c1}\includegraphics[width=3.9cm,height=3.8cm]{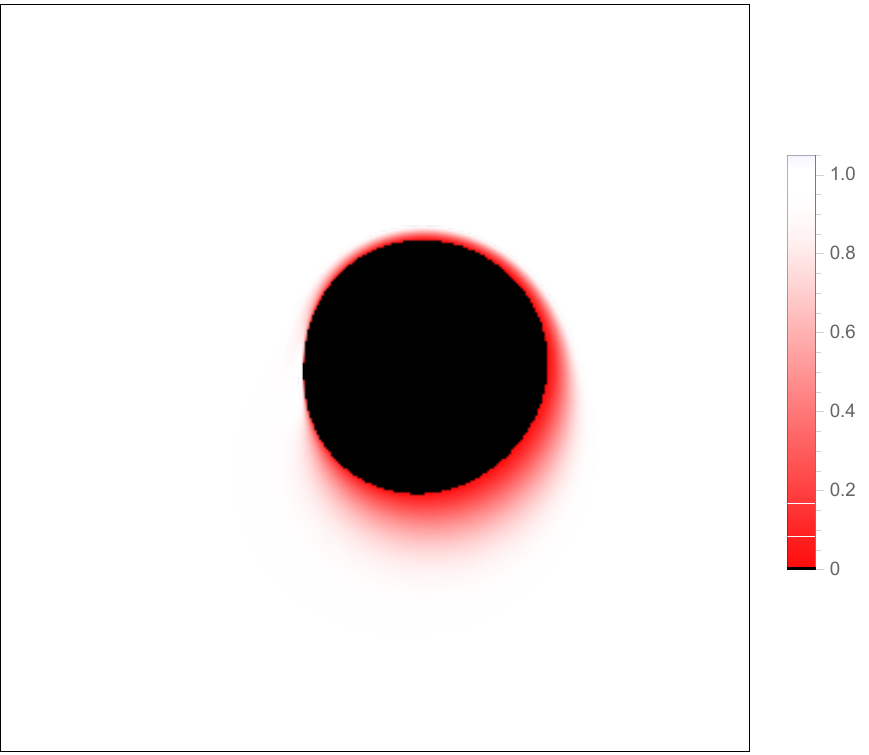}}
\subfigure[\tiny][$R_{s}=0.1,~\rho_{c}=0.65$]{\label{d1}\includegraphics[width=3.9cm,height=3.8cm]{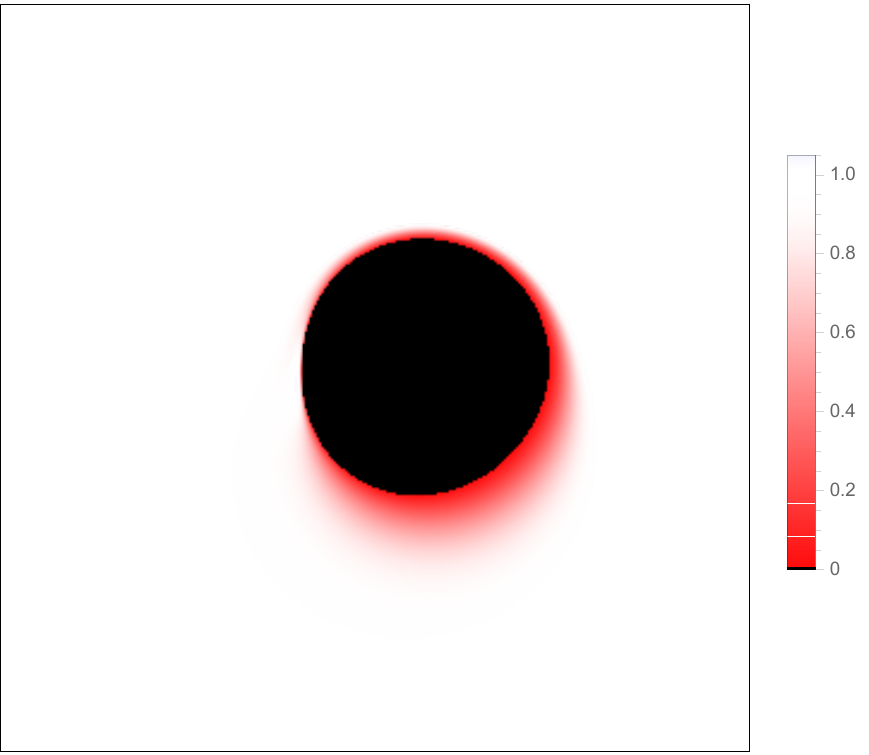}}
\subfigure[\tiny][$R_{s}=0.2,~\rho_{c}=0.05$]{\label{a2}\includegraphics[width=3.9cm,height=3.8cm]{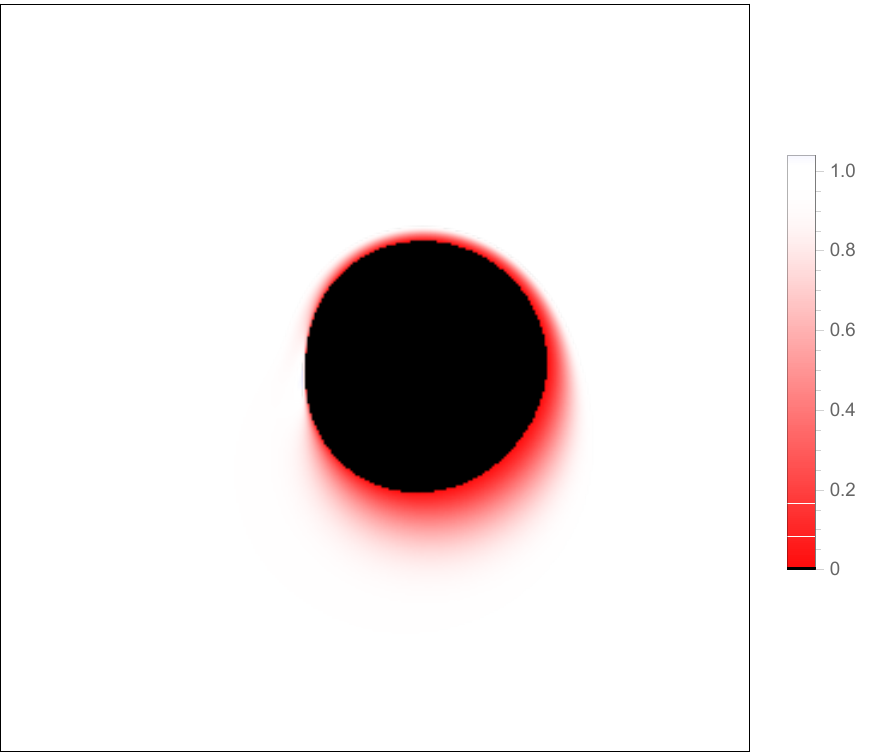}}
\subfigure[\tiny][$R_{s}=0.2,~\rho_{c}=0.25$]{\label{b2}\includegraphics[width=3.9cm,height=3.8cm]{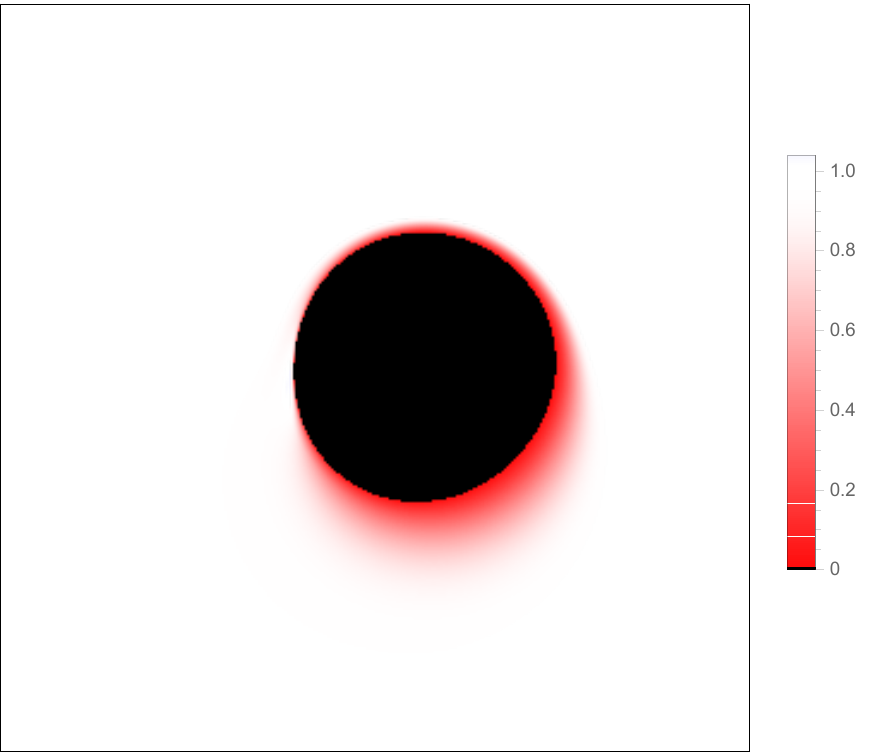}}
\subfigure[\tiny][$R_{s}=0.2,~\rho_{c}=0.45$]{\label{d2}\includegraphics[width=3.9cm,height=3.8cm]{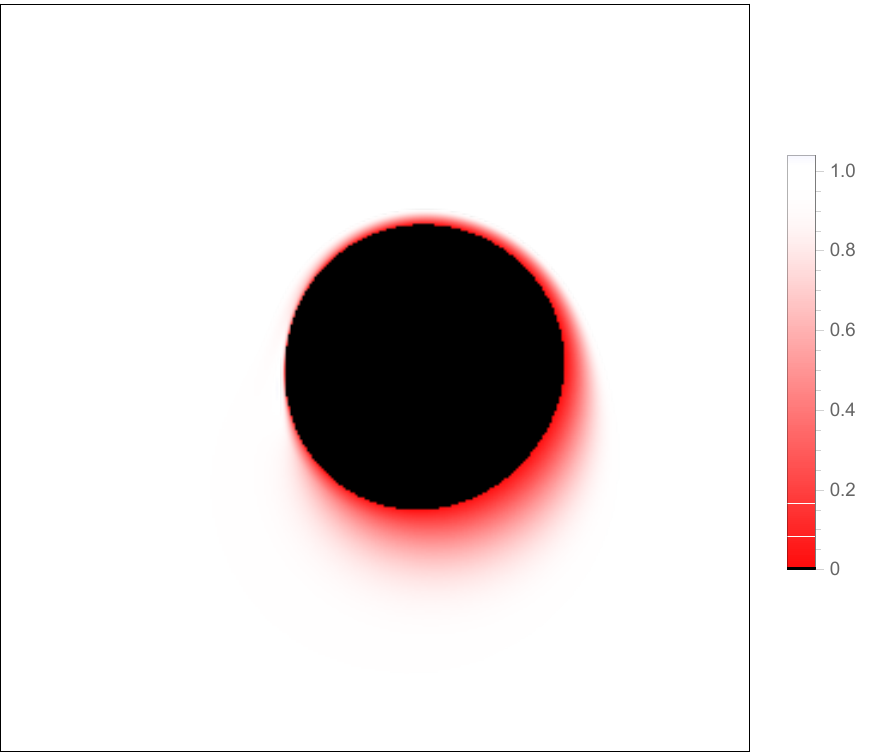}}
\subfigure[\tiny][$R_{s}=0.2,~\rho_{c}=0.65$]{\label{d2}\includegraphics[width=3.9cm,height=3.8cm]{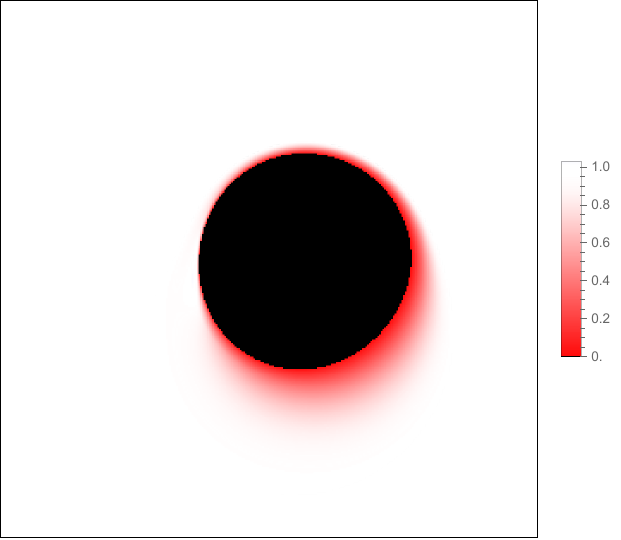}}
\subfigure[\tiny][$R_{s}=0.3,~\rho_{c}=0.05$]{\label{a3}\includegraphics[width=3.9cm,height=3.8cm]{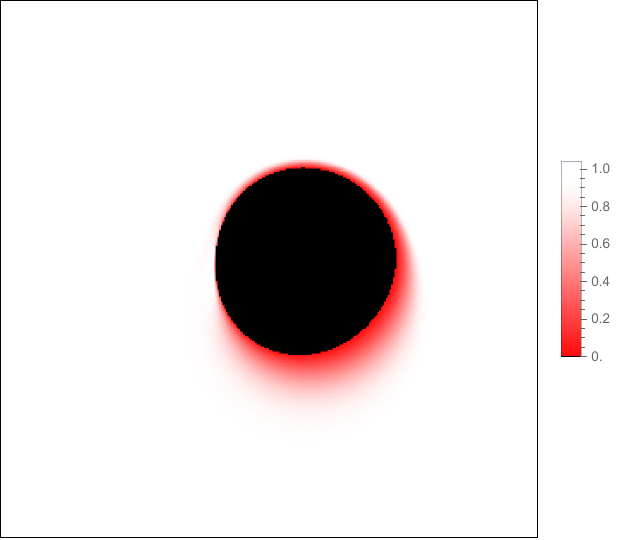}}
\subfigure[\tiny][$R_{s}=0.3,~\rho_{c}=0.25$]{\label{b3}\includegraphics[width=3.9cm,height=3.8cm]{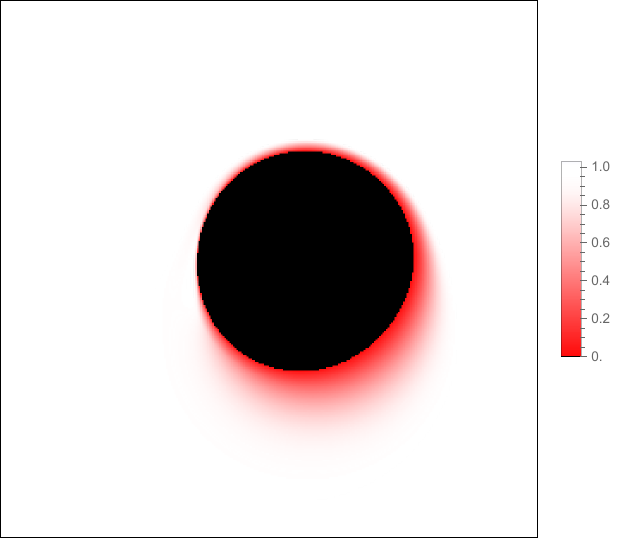}}
\subfigure[\tiny][$R_{s}=0.3,~\rho_{c}=0.45$]{\label{c3}\includegraphics[width=3.9cm,height=3.8cm]{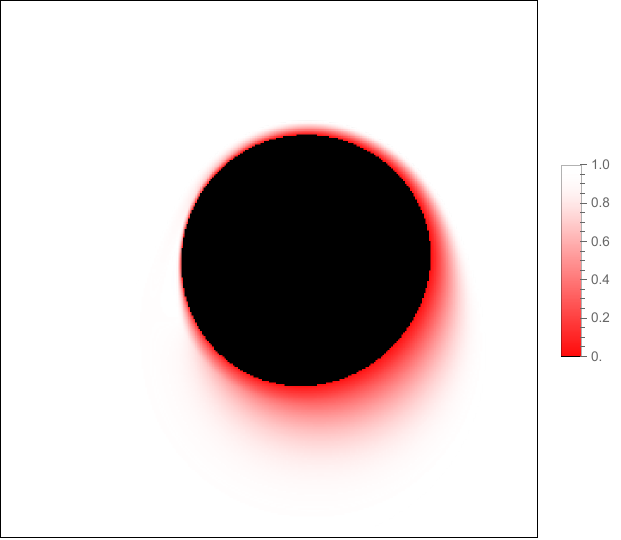}}
\subfigure[\tiny][$R_{s}=0.3,~\rho_{c}=0.65$]{\label{d3}\includegraphics[width=3.9cm,height=3.8cm]{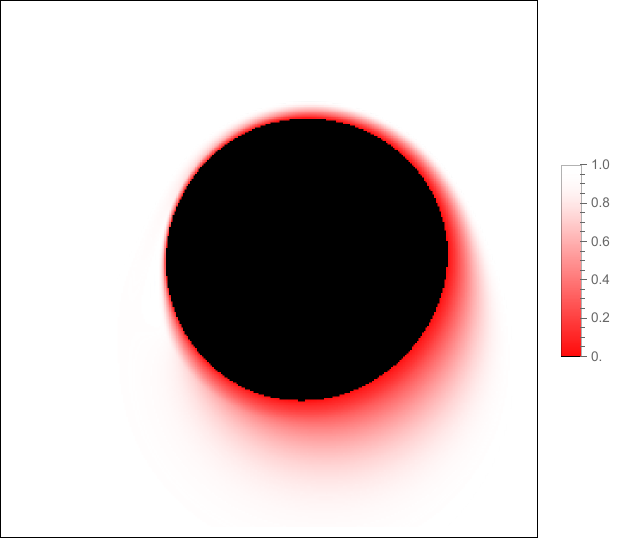}}
\caption{The red-shifts configuration of lensed images with prograde
flow under several values of $\rho_{c}$ and $R_{s}$ with $a=0.998$
and $\theta_{obs}=60^{\circ}$. Further, the horizontal and vertical
axis correspond to $x/M$ and $y/M$, respectively.}\label{sf7}
\end{center}
\end{figure}

In Fig. \textbf{\ref{sf6}}, the red-shift configuration of the
direct images are interpreted under the same set of model parameters
values as defined in Fig. \textbf{\ref{sf5}}. To visually show the
red-shift factor, a constant linear colour pattern is used, where
blue and red colours correspond to blue-shift and red-shift factors.
In each panel, the central black region reveals the inner shadow
cast by the accretion disk, which is surrounded by the prolongation
of the event horizon of the BH. From the upper row of Fig.
\textbf{\ref{sf6}}, it is observed that when $R_{s}$ is fixed and
$\rho_{c}$ varies from left to right, both distributions of
blue-shift and red-shift appear on the left and right side of
the screen, respectively. Particularly, the red colour map is
distributed around the inner shadow of the image, and it is spread in
more space on the screen as compared to the blue-shift. When the value
of scale radius increases such as $R_{s}=0.2$ (see second row of
Fig. \textbf{\ref{sf6}}), the red-shift factor is more obvious on
the screen, while the optical appearance of blue-shift is slightly
dim. Moreover, the modification of both $R_{s}$ and $\rho_{c}$
exerts a considerable effect on the central region, which is
slightly increased with the increment of $\rho_{c}$. Subsequently,
when we further increase the value of critical radius $R_{s}=0.3$,
one can observe that the range of red-shift factor is simultaneously
expanding on the screen as compared to blue-shift. Another notable
phenomenon is observed, the augmentation of $R_{s}$ leads to increase
the central dark region along with the deformation of $\rho_{c}$,
which is more obvious as compared to previous ones. In addition, we
calculated the maximal blue-shift $g_{max}$ for different values of
$R_{s}$ and $\rho_{c}$ in Table \textbf{\ref{tab1}}. From this
Table, one can see that the numerical values of $g_{max}$ decrease
from top to bottom when $R_{s}=0.1,~0.2,~0.3,~0.4$ and $0.7$, while
it increase when $R_{s}=0.5$ and $0.6$. On the other hand, the
maximal blue-shift $g_{max}$ decreases from left to right for
$\rho_{c}=0.05,~0.15,~0.55$ and $0.65$. The optical appearance of
the red-shift distribution for the lensed images with prograde flow
is illustrated in Fig. \textbf{\ref{sf7}}, where the parameter
values correspond to those in Fig. \textbf{\ref{sf6}}.
In all cases, it is observed that the outer boundary of the inner
shadow is enveloped by a strict red ring, which happens due to
the radiation of light particles within the falling region,
producing a prominent exhibition of a red-shift. Consequently, with
the variations of parameters, the lensed images of the accretion disk
predominantly interpret the red-shift features, while the blue-shift
factor is notably diminished. Particularly, the influence of
blue-shift is not observed in lensed images. However, with the
enhancement of the scale radius $R_{s}$ and critical density
$\rho_{c}$, area of the central dark region increases significantly,
while the phenomenon of red-shift is suppressed from top to bottom.
Moreover, the visual appearance of the red-shift colour map is
spread in the lower right quadrant of the screen.
\begin{figure}[H]
\begin{center}
\subfigure[\tiny][$R_{s}=0.1,~\rho_{c}=0.05$]{\label{a1}\includegraphics[width=3.9cm,height=3.8cm]{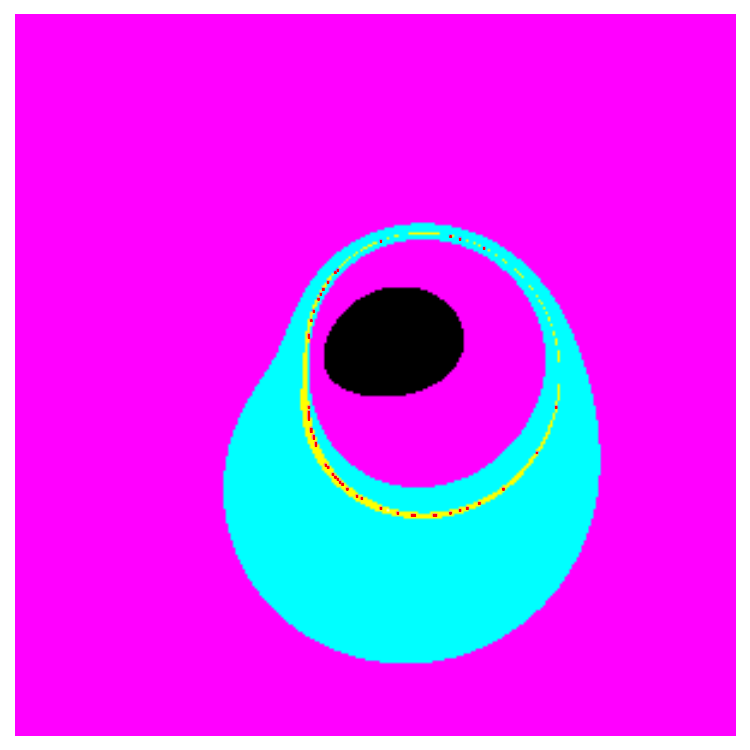}}
\subfigure[\tiny][$R_{s}=0.1,~\rho_{c}=0.25$]{\label{b1}\includegraphics[width=3.9cm,height=3.8cm]{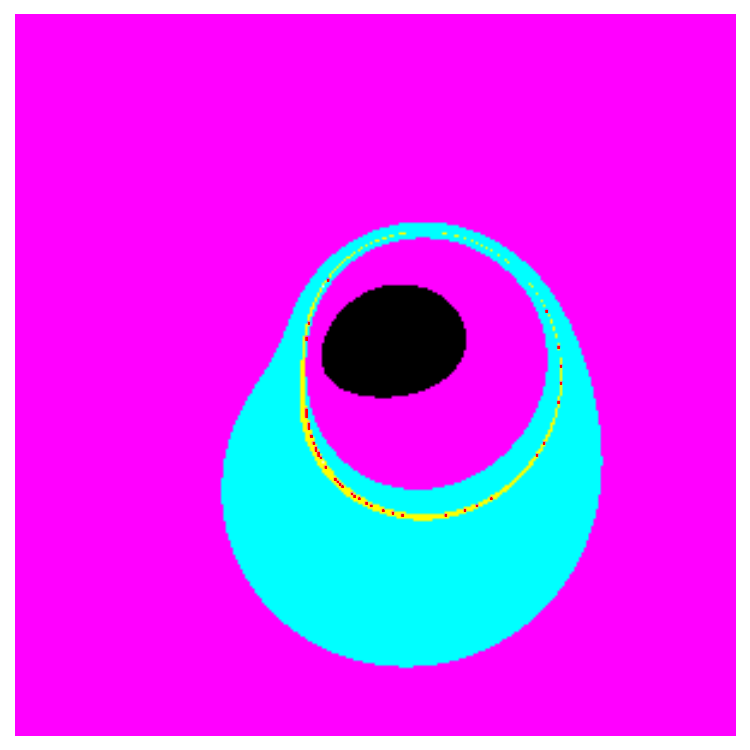}}
\subfigure[\tiny][$R_{s}=0.1,~\rho_{c}=0.45$]{\label{c1}\includegraphics[width=3.9cm,height=3.8cm]{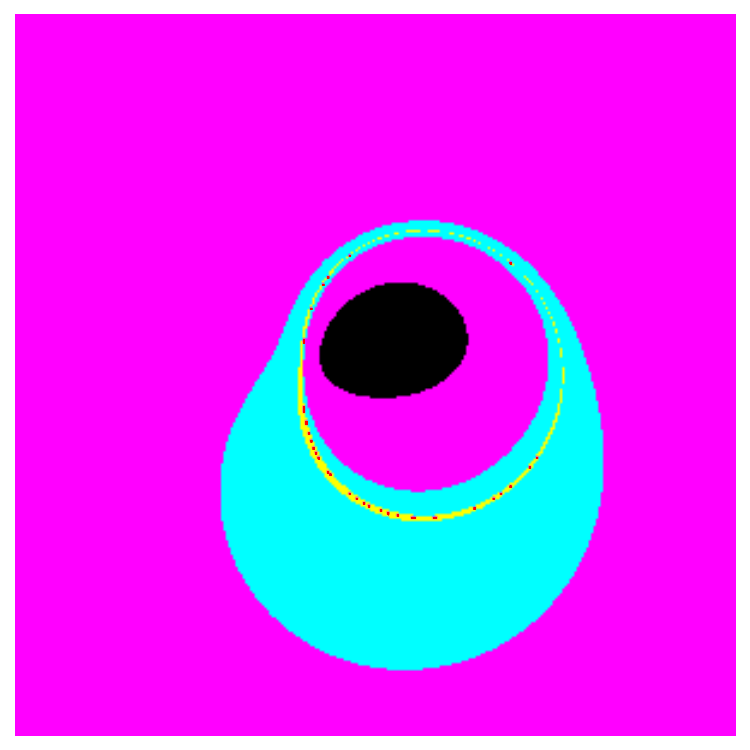}}
\subfigure[\tiny][$R_{s}=0.1,~\rho_{c}=0.65$]{\label{d1}\includegraphics[width=3.9cm,height=3.8cm]{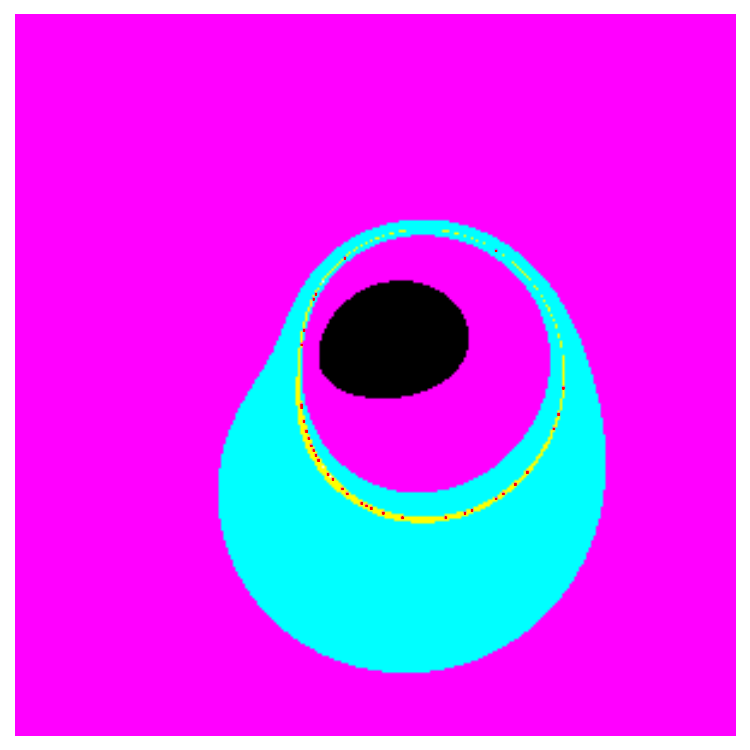}}
\subfigure[\tiny][$R_{s}=0.2,~\rho_{c}=0.05$]{\label{a2}\includegraphics[width=3.9cm,height=3.8cm]{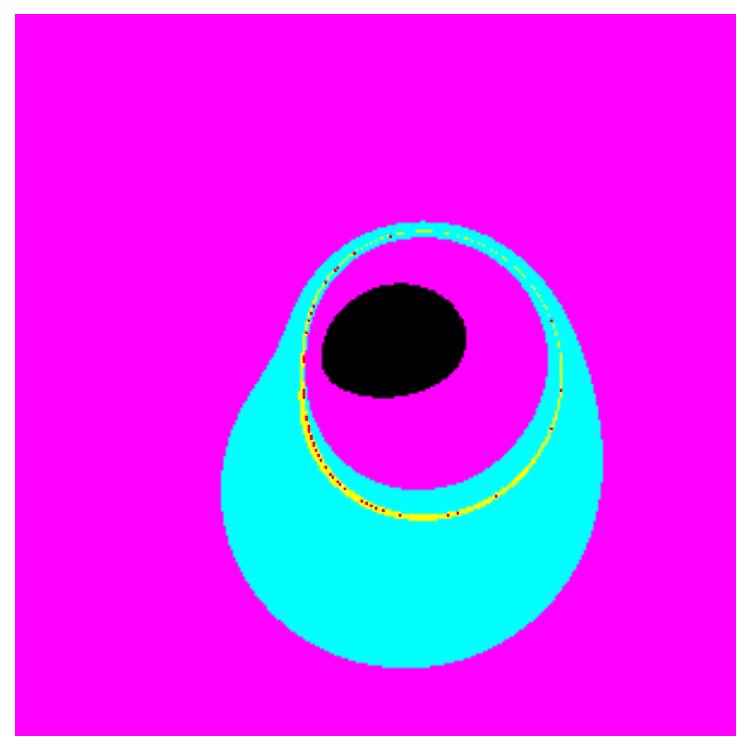}}
\subfigure[\tiny][$R_{s}=0.2,~\rho_{c}=0.25$]{\label{b2}\includegraphics[width=3.9cm,height=3.8cm]{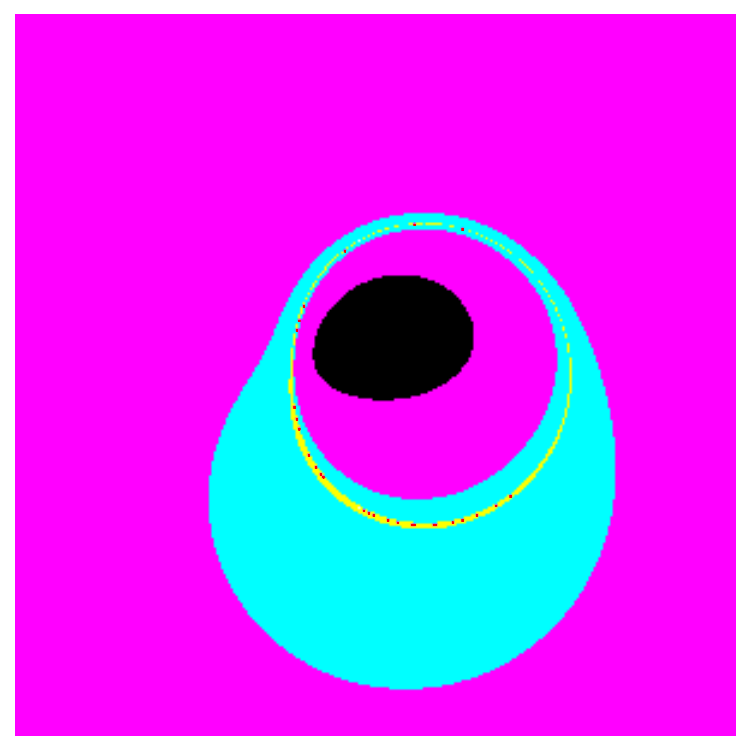}}
\subfigure[\tiny][$R_{s}=0.2,~\rho_{c}=0.45$]{\label{d2}\includegraphics[width=3.9cm,height=3.8cm]{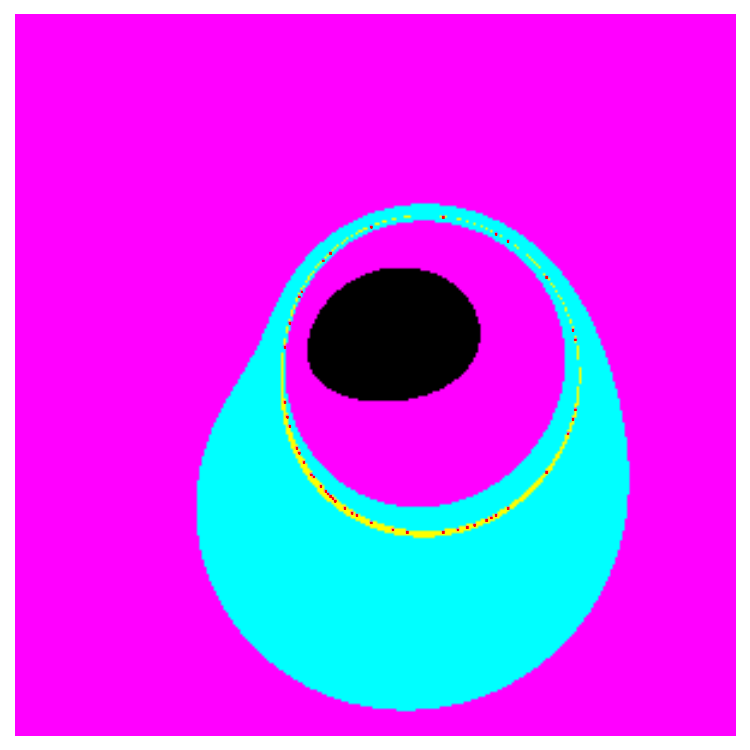}}
\subfigure[\tiny][$R_{s}=0.2,~\rho_{c}=0.65$]{\label{d2}\includegraphics[width=3.9cm,height=3.8cm]{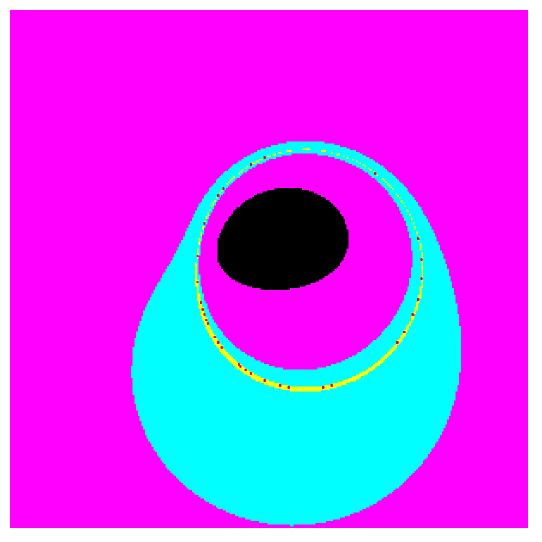}}
\subfigure[\tiny][$R_{s}=0.3,~\rho_{c}=0.05$]{\label{a3}\includegraphics[width=3.9cm,height=3.8cm]{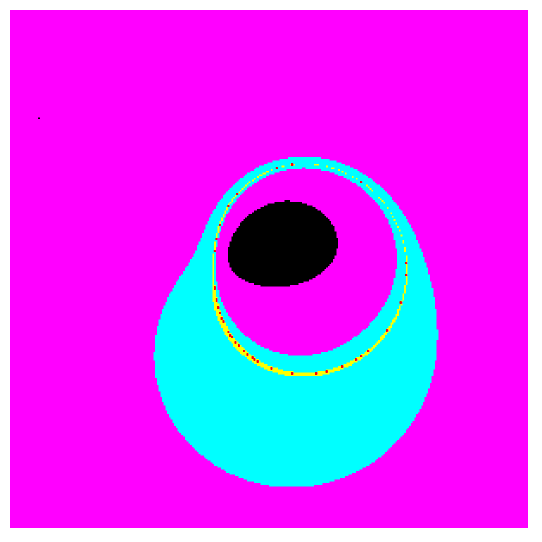}}
\subfigure[\tiny][$R_{s}=0.3,~\rho_{c}=0.25$]{\label{b3}\includegraphics[width=3.9cm,height=3.8cm]{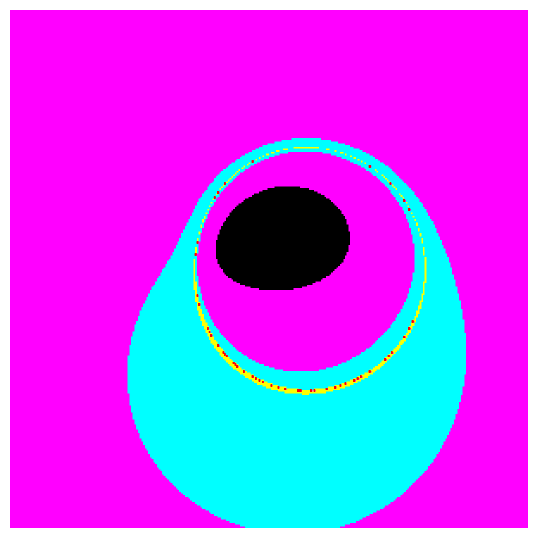}}
\subfigure[\tiny][$R_{s}=0.3,~\rho_{c}=0.45$]{\label{c3}\includegraphics[width=3.9cm,height=3.8cm]{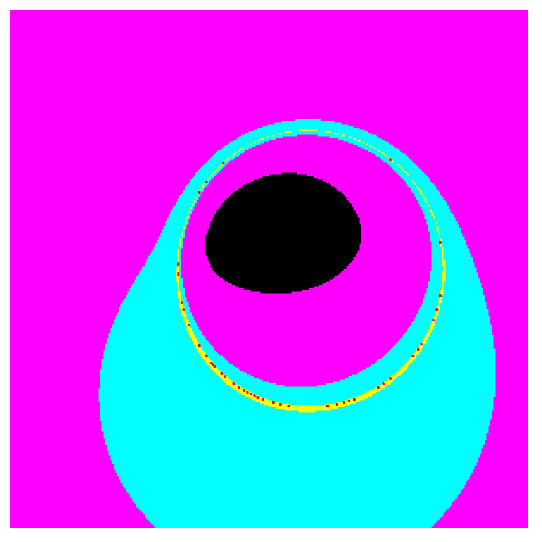}}
\subfigure[\tiny][$R_{s}=0.3,~\rho_{c}=0.65$]{\label{d3}\includegraphics[width=3.9cm,height=3.8cm]{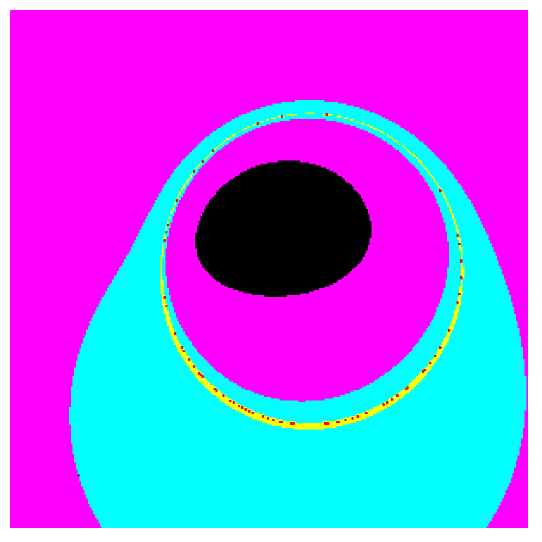}}
\caption{The lensing bands of rotating BH in the presence of CDM
halo with prograde flow under several values of $\rho_{c}$ and
$R_{s}$ with $a=0.998$ and $\theta_{obs}=60^{\circ}$. The colours
magenta, cyan and yellow correspond to the direct, lensed and photon
ring images, respectively. Further, the horizontal and vertical axis
correspond to $x/M$ and $y/M$, respectively.}\label{sf8}
\end{center}
\end{figure}
All these results imply that the size of the inner shadow as well as
the visual appearance of the red-shift is closely related to the
values of associated parameters. In Table \textbf{\ref{tab2}}, we
calculated the numerical values of maximal blue-shift $g_{max}$ for
different values of $R_{s}$ and $\rho_{c}$ of lensed images. From
this table, one can observe that there is a minor impact of $R_{s}$
and $\rho_{c}$ on the blue-shift distribution, which is almost
invisible on the observer's screen.

\begin{table}[H]\centering
\begin{tabular}{|c|c|c|c|c|c|c|c|}
\hline \diagbox{$\rho_c$}{$R_s$} & 0.1 & 0.2 & 0.3 & 0.4 & 0.5 & 0.6
& 0.7 \\ \hline 0.05 & 1.03478 & 1.0449 & 1.0335 & 1.03854 & 1.01931
& 1.0189 & 1.0215 \\ \hline 0.15 & 1.02189 & 1.01886 & 1.01142 &
1.02802 & 1.01287 & 1.01308 & 1.00808 \\ \hline 0.25 & 1.03771 &
1.01877 & 1.02535 & 1.00677 & 1.01305 & 1.00343 & 1
\\ \hline 0.35 & 1.04602 & 1.02229 & 1.02762 & 1.01973 & 1.00802 &
1.0057 & 0.746386 \\ \hline 0.45 & 1.02398 & 1.0221 & 1.00708 &
1.00794 & 1.0055 & 0.967922 & 0.016851 \\ \hline 0.55 & 1.02257 &
1.01958 & 1.02489 & 1.01631 & 1.00815 & 0.334845 & 0 \\ \hline 0.65
& 1.03841 & 1.01647 & 1.01064 & 1.00634 & 1 & 0.0735375 & 0 \\
\hline
\end{tabular}
\caption{The maximal blue-shift $g_{max} $ of lensed images under
different values of $R_s$ and $\rho_c$ with $a=0.998$ and
$\theta_{obs}=60^{\circ}$.}\label{tab2}
\end{table}

For better understanding of the differences between direct and
lensed images, we interpret their observed fluxes for different
values of relevant parameters in Fig. \textbf{\ref{sf8}}, where the
assigned values of relevant parameters correspond to those in Fig.
\textbf{\ref{sf7}}. Upon comparison, from the first row of Fig.
\textbf{\ref{sf8}}, we observed that the cyan bands always appear in
the lower half quadrant of the screen, and slightly move towards the
left side with the enhancement of the critical density $\rho_{c}$.
Moreover, the lens bands are slightly deformed towards the lower
side of the screen, while the yellow circular structure, which
corresponds to the photon ring remains the same in all cases. When the
scale radius $R_{s}=0.2$ (see second row of Fig.
\textbf{\ref{sf8}}), variations in the $\rho_{c}$ significantly
deformed the lensed bands in the lower half of the screen,
while the radius of photon ring slightly increases with the aid of
$\rho_{c}$. When we further increase the value of $R_{s}$ such as
$R_{s}=0.3$, the optical appearance of direct and lensed bands is
deformed prominent, and shows the same effect as discussed in the
previous cases. Particularly, when $R_{s}=0.3$ and $\rho_{c}=0.65$,
the lensed bands occupied more space on the screen compared to
previous ones. In all these cases, the photon ring always lies
exactly within the confines of magenta and cyan bands. Further, the
inner shadow exhibits a hat-like shape, which is more obvious for
larger values of both $R_{s}$ and $\rho_{c}$.

\begin{figure}[H]
\begin{center}
\subfigure[\tiny][$R_{s}=0.2,~\rho_{c}=0.05$]{\label{a1}\includegraphics[width=3.9cm,height=3.8cm]{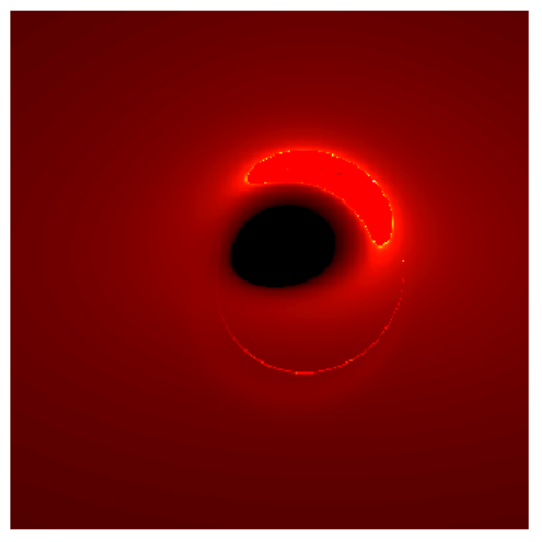}}
\subfigure[\tiny][$R_{s}=0.2,~\rho_{c}=0.25$]{\label{b1}\includegraphics[width=3.9cm,height=3.8cm]{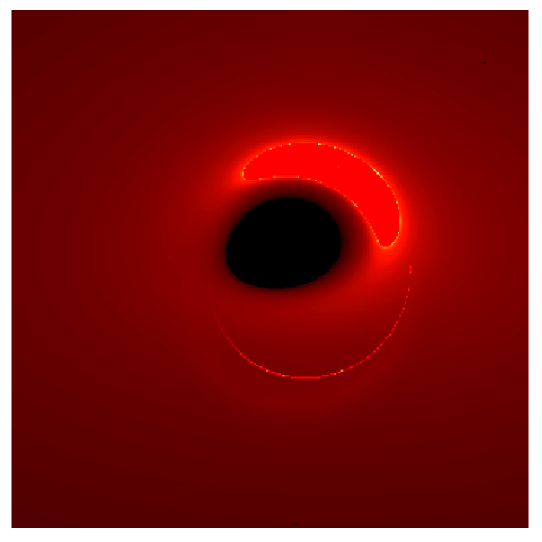}}
\subfigure[\tiny][$R_{s}=0.2,~\rho_{c}=0.45$]{\label{c1}\includegraphics[width=3.9cm,height=3.8cm]{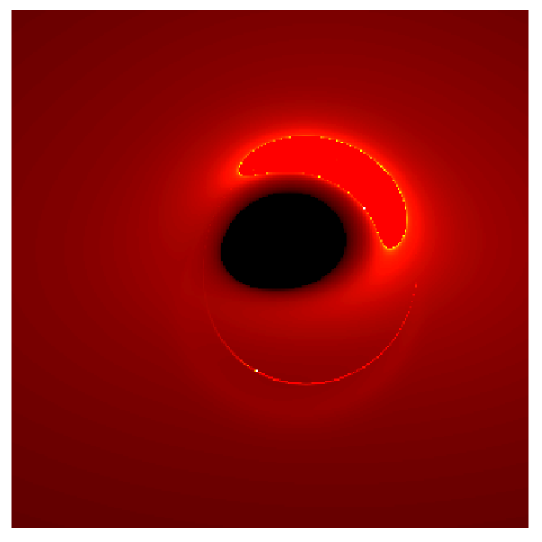}}
\subfigure[\tiny][$R_{s}=0.2,~\rho_{c}=0.65$]{\label{d1}\includegraphics[width=3.9cm,height=3.8cm]{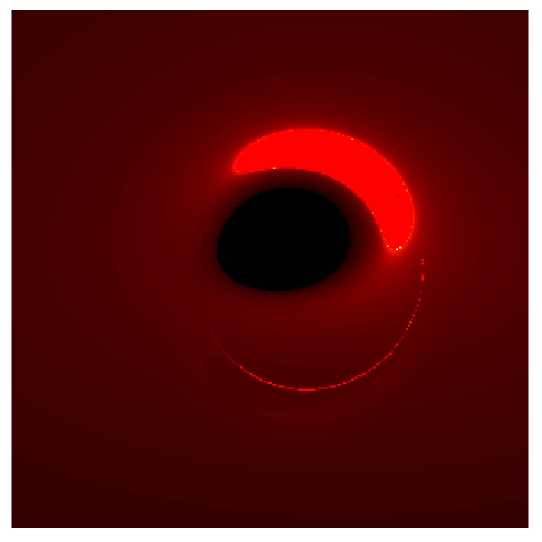}}
\caption{Shadows cast by rotating BH in the presence of CDM halo
surrounded by retrograde flow at $230$ GHz under several values of
$\rho_{c}$ with $R_{s}=0.2$,~$a=0.998$ and
$\theta_{obs}=60^{\circ}$. Further, the horizontal and vertical axis
correspond to $x/M$ and $y/M$, respectively.}\label{sf9}
\end{center}
\end{figure}

Now, we discuss the optical properties of BH shadows under the
retrograde accretion flow observed in the observer's frame. In Fig.
\textbf{\ref{sf9}}, we depict the influence of variation in critical
density $\rho_{c}$ on the geometrical shape of a rotating BH with
CDM halo, which is enveloped by a retrograde accretion flow.
Looking at Figs. \textbf{\ref{sf9}} (a-d), one finds that the inner
shadow region gradually increases with the variation of $\rho_{c}$,
however the position of the Einstein ring remains almost the same in
all cases. In comparison to the Fig. \textbf{\ref{sf5}} (second
row), the visual appearance of the BH image is significantly
depressed due to the gravitational red-shift. Moreover, in this
case, the optical depth reduced the visibility of the lensed image
as it appeared a crescent or eyebrow-shaped bright region on the
upper half of the screen. This is because of the jet material
from the side of the event horizon exhibits low brightness, and
during the imaging process, the radiating material helped to
increase the total optical depth, which is confined in the
equatorial plane. Hence, the direct geometrical impact of the
equatorial emission being confined at the event horizon is still
discernible, even if it is hardly perceptible.

\begin{figure}[H]
\begin{center}
\subfigure[\tiny][$R_{s}=0.2,~\rho_{c}=0.05$]{\label{a1}\includegraphics[width=3.9cm,height=3.8cm]{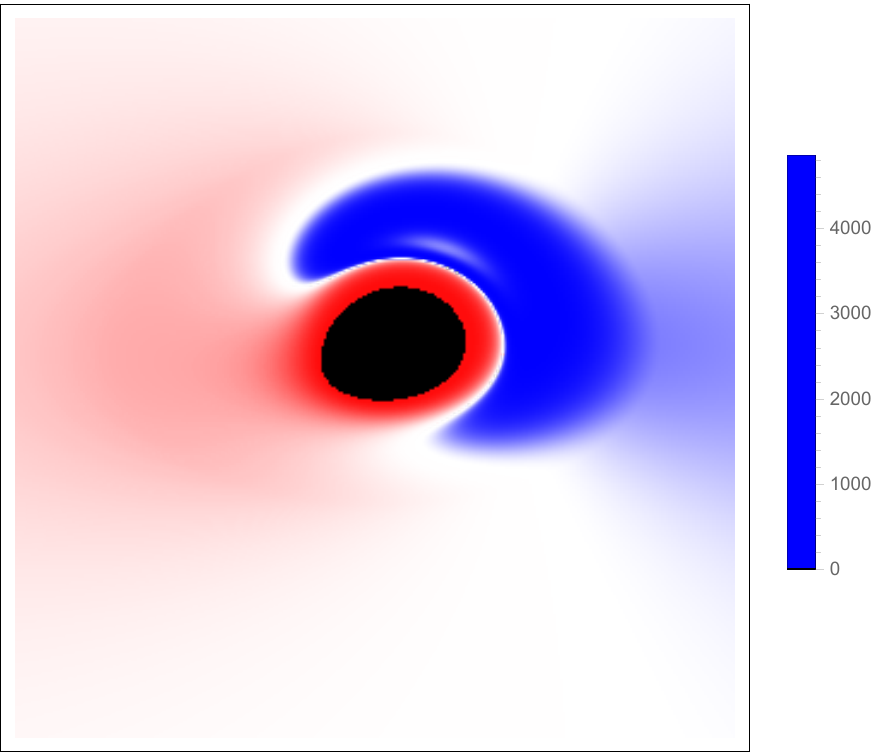}}
\subfigure[\tiny][$R_{s}=0.2,~\rho_{c}=0.25$]{\label{b1}\includegraphics[width=3.9cm,height=3.8cm]{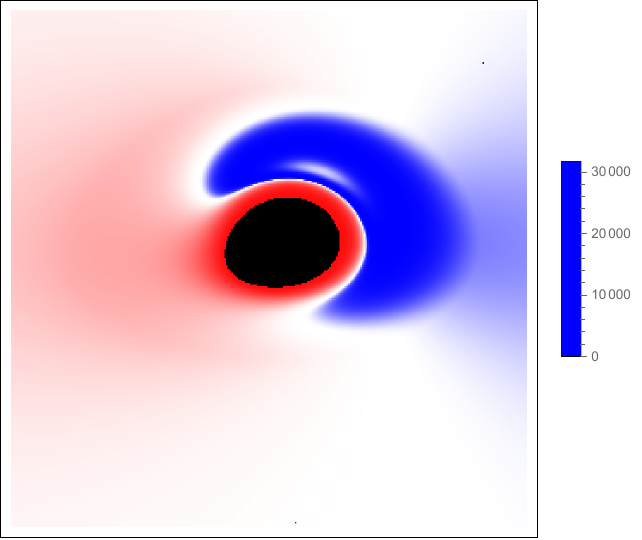}}
\subfigure[\tiny][$R_{s}=0.2,~\rho_{c}=0.45$]{\label{c1}\includegraphics[width=3.9cm,height=3.8cm]{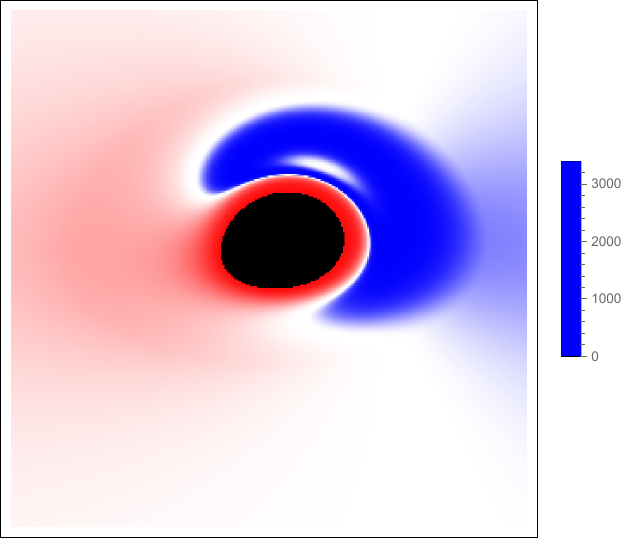}}
\subfigure[\tiny][$R_{s}=0.2,~\rho_{c}=0.65$]{\label{d1}\includegraphics[width=3.9cm,height=3.8cm]{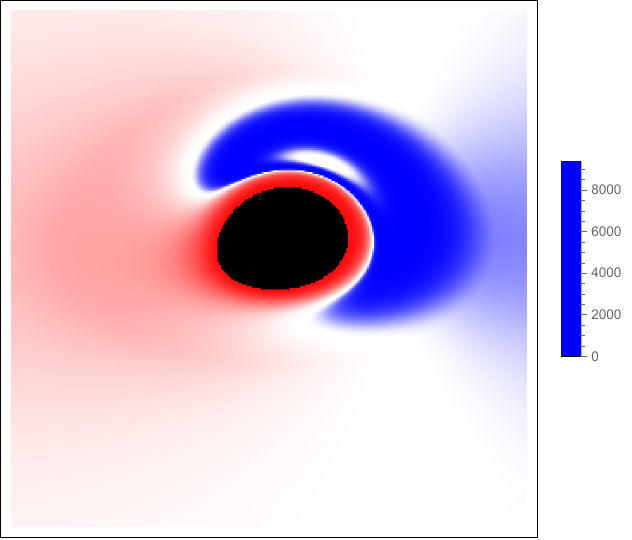}}
\caption{The red-shifts configuration of direct images with
retrograde flow under several values of $\rho_{c}$ with
$R_{s}=0.2$,~$a=0.998$ and $\theta_{obs}=60^{\circ}$. Further, the
horizontal and vertical axis correspond to $x/M$ and $y/M$,
respectively.}\label{sf10}
\end{center}
\end{figure}
\begin{figure}[H]
\begin{center}
\subfigure[\tiny][$R_{s}=0.2,~\rho_{c}=0.05$]{\label{a1}\includegraphics[width=3.9cm,height=3.8cm]{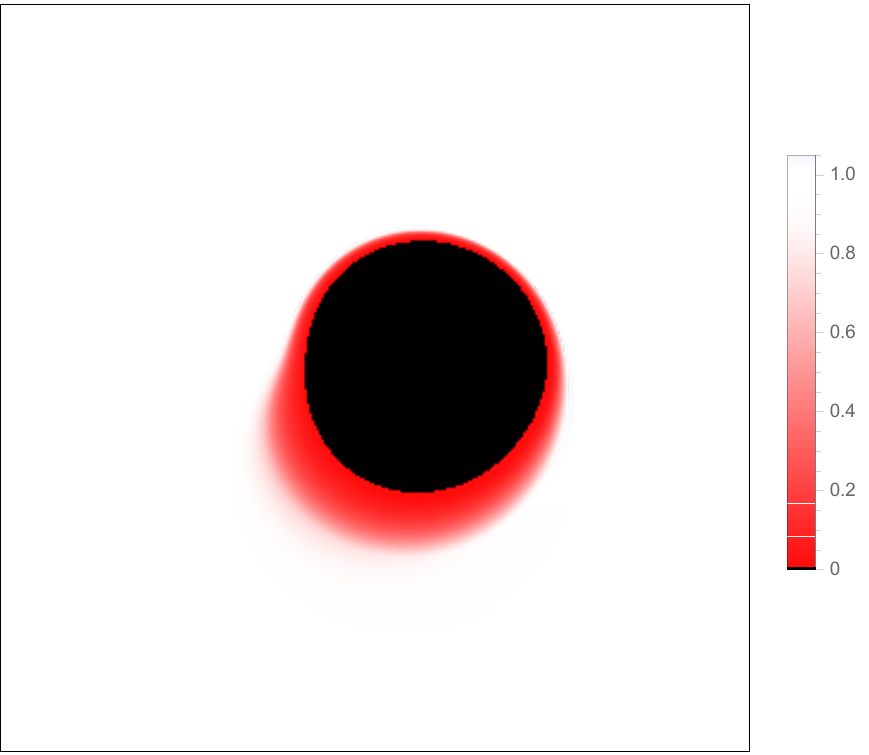}}
\subfigure[\tiny][$R_{s}=0.2,~\rho_{c}=0.25$]{\label{b1}\includegraphics[width=3.9cm,height=3.8cm]{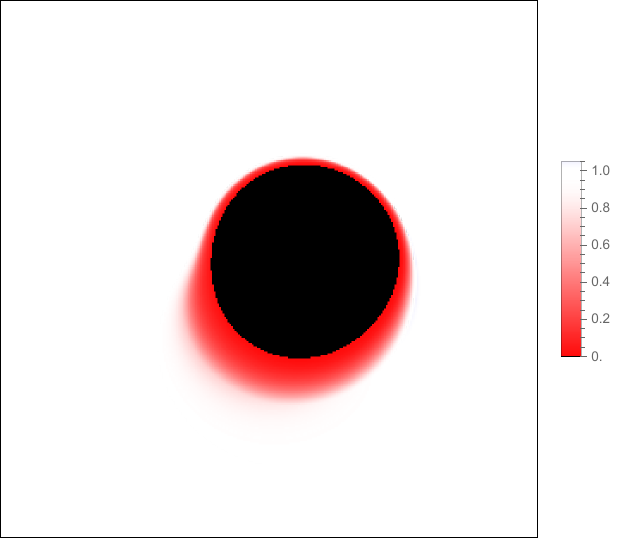}}
\subfigure[\tiny][$R_{s}=0.2,~\rho_{c}=0.45$]{\label{c1}\includegraphics[width=3.9cm,height=3.8cm]{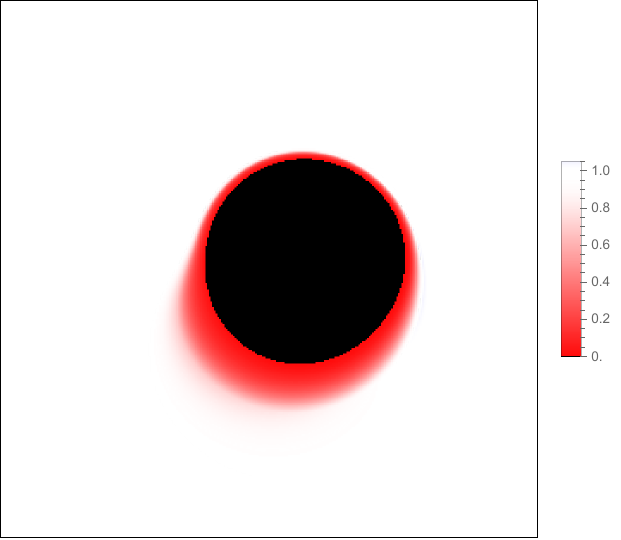}}
\subfigure[\tiny][$R_{s}=0.2,~\rho_{c}=0.65$]{\label{d1}\includegraphics[width=3.9cm,height=3.8cm]{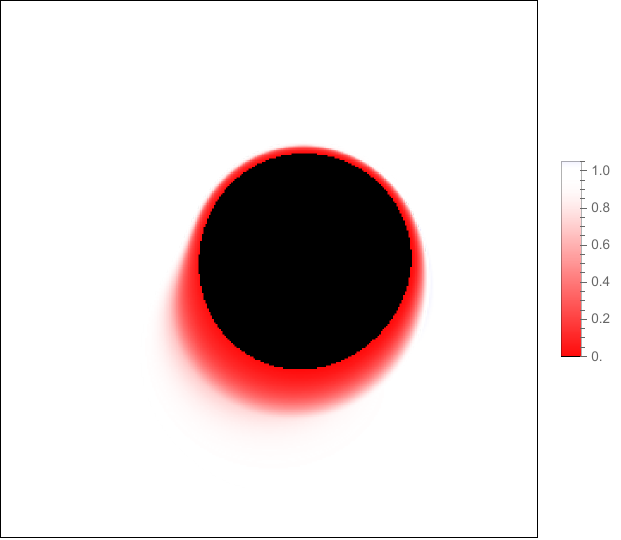}}
\caption{The red-shifts configuration of lensed images with
retrograde flow under several values of $\rho_{c}$ with
$R_{s}=0.2$,~$a=0.998$ and $\theta_{obs}=60^{\circ}$. Further, the
horizontal and vertical axis correspond to $x/M$ and $y/M$,
respectively.}\label{sf11}
\end{center}
\end{figure}

In order to see the optical signatures of red-shifts configuration
of direct images with the direction of retrograde flow, we plot the
physical interpretation of these distributions in Fig.
\textbf{\ref{sf10}}. From these images, one can see that as the
value of critical density $\rho_{c}$ increases, the red-shift
lensing phenomenon expands in more space and envelops the
blue-shift zone more obviously. The reason for this phenomenon is due
to the motion of retrograde flow and relativistic jets, which are
essentially placed along the equatorial plane, where the flow of
radiated particles starts. In the case of lensed images, we observed
that only the red-shift distribution displays on the lower left
quadrant of the screen, and its strip gradually expands with the aid
of $\rho_{c}$, see Fig. \textbf{\ref{sf11}}.

\begin{figure}[H]
\begin{center}
\subfigure[\tiny][$R_{s}=0.2,~\rho_{c}=0.05$]{\label{a1}\includegraphics[width=3.9cm,height=3.8cm]{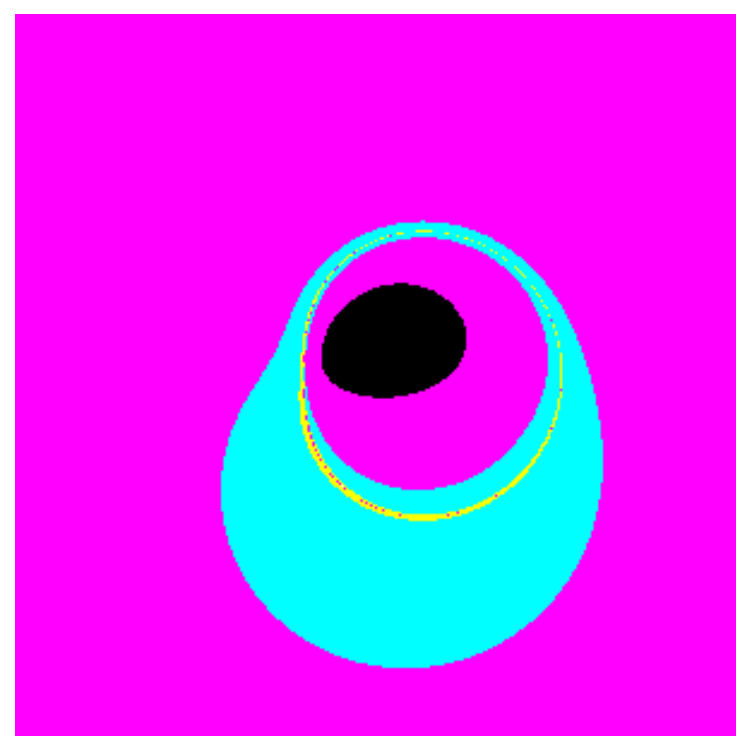}}
\subfigure[\tiny][$R_{s}=0.2,~\rho_{c}=0.25$]{\label{b1}\includegraphics[width=3.9cm,height=3.8cm]{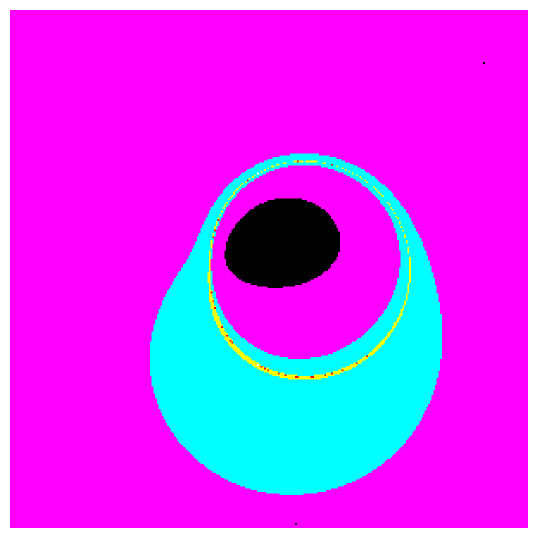}}
\subfigure[\tiny][$R_{s}=0.2,~\rho_{c}=0.45$]{\label{c1}\includegraphics[width=3.9cm,height=3.8cm]{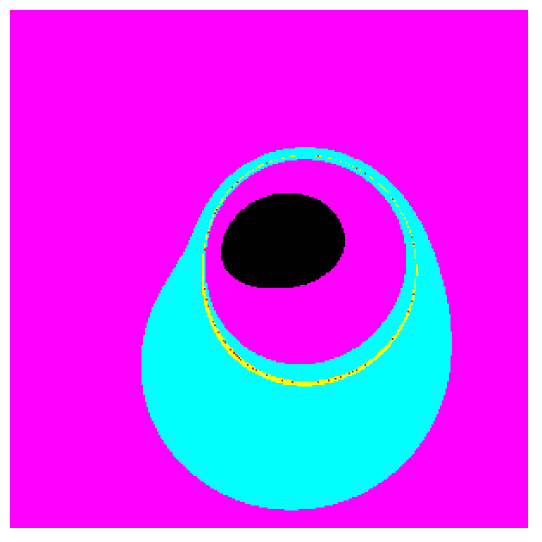}}
\subfigure[\tiny][$R_{s}=0.2,~\rho_{c}=0.65$]{\label{d1}\includegraphics[width=3.9cm,height=3.8cm]{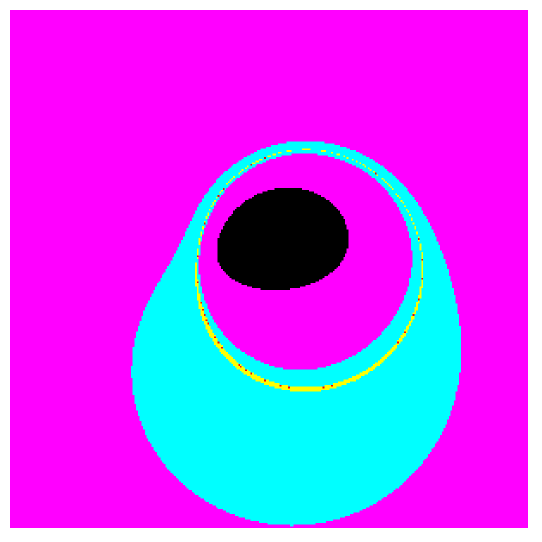}}
\caption{The lensing bands of rotating BH in the presence of CDM
halo with retrograde flow under several values of $\rho_{c}$ with
$R_{s}=0.2$,~$a=0.998$ and $\theta_{obs}=60^{\circ}$. The colours
magenta, cyan and yellow correspond to the direct, lensed and photon
ring images, respectively. Further, the horizontal and vertical axis
correspond to $x/M$ and $y/M$, respectively.}\label{sf12}
\end{center}
\end{figure}

Actually, the emission from the side of accretion disk is
somewhat smaller compared to the inner shadow region. This happens
because the radiation arises from the place that crossed the event
horizon at their highest latitude inside the equatorial region. In
Fig. \textbf{\ref{sf12}}, we interpret the lensing bands of the
rotating BH in the presence of CDM halo for different values of
$\rho_{c}$ with retrograde accretion flow. In contrast, it is
observed that the lensing bands are gradually expanding towards the
lower left side of the screen. In all cases, the inner shadow
exhibits a hat-like shape, which varies with the variations of
$\rho_{c}$. Moreover, the yellow bands (photon ring) always lie
within the confines of magenta and cyan colours, and the size of the
photon ring slightly increases towards the lower side of the screen
with the increasing values of $\rho_{c}$.

\subsection{Comparison with EHT Results}

Now, we are going to discuss the constraints on the relevant
Parameters of BH, using the recent observational bounds from the EHT data.
These findings of M$87^{\ast}$ and Sgr $A^{\ast}$ will help us to
evaluate the constraints on the critical density and the scale
radius of these BHs. We will compare the observational bounds
obtained for these BHs to gain insights into how CDM halo might
affect the characteristics of rotating BHs. For desired results, we
calculate the angular diameter of BH shadows to construct a
comparative analysis with the angular diameter of M$87^{\ast}$ and
Sgr $A^{\ast}$.

\begin{figure}[H]
\begin{center}
\subfigure[\tiny][$R_{s}=a=0.22$]{\label{c1}\includegraphics[width=8cm,height=6cm]{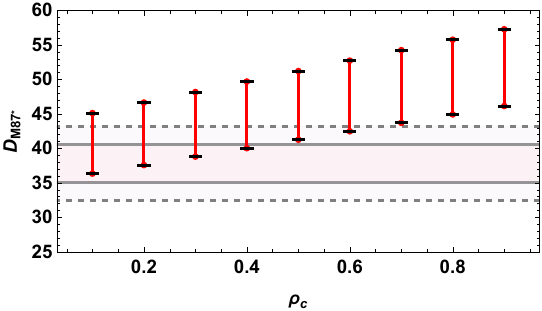}}
\subfigure[\tiny][$\rho_{c}=a=0.3$]{\label{d1}\includegraphics[width=8cm,height=6cm]{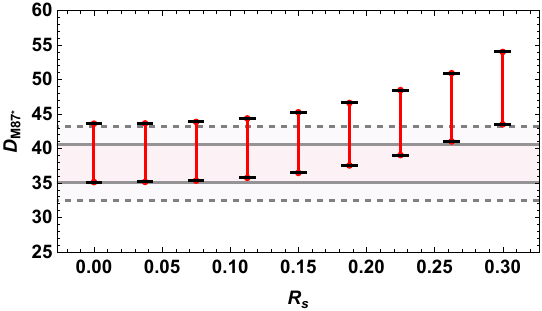}}
\subfigure[\tiny][$R_s=a=0.22$]{\label{a1}\includegraphics[width=8cm,height=6cm]{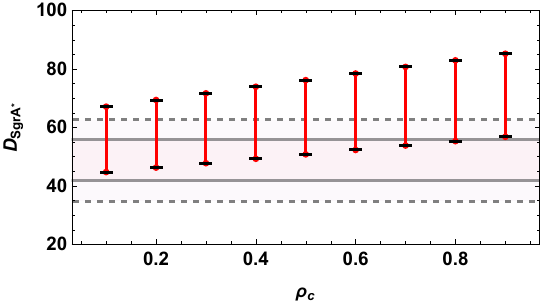}}
\subfigure[\tiny][$\rho_{c}=a=0.3$]{\label{b1}\includegraphics[width=8cm,height=6cm]{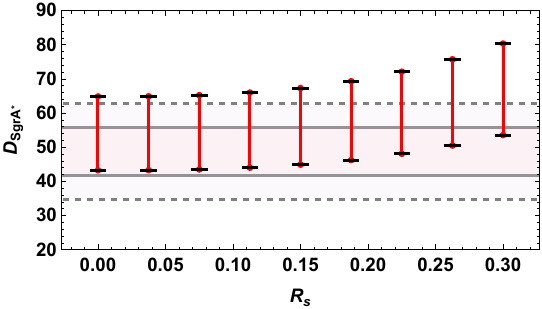}}
\caption{Comparison of shadow angular diameter $D$ for space-time
(\ref{s2}) with EHT results for M$87^{\ast}$ (top row) and Sgr
$A^{\ast}$ (bottom row). The solid and dashed grey lines indicate
the $1\sigma$ and $2\sigma$ confidence levels for $D$, respectively.
The red segments represent the approximated intervals with endpoints
indicated by thickened black tick marks.}\label{israr1}
\end{center}
\end{figure}

Corresponding to such constraints on the relevant parameters, the BH
is assumed to mimic either M$87^{\ast}$ or Sgr $A^{\ast}$, if the
angular diameter of the BH shadow lies within the $1\sigma$ and
$2\sigma$ confidence level. As outlined in \cite{sd35,sd65}, the
angular diameter of the BH is defined as
$D=2\hat{R}_{d}\frac{\mathcal{M}}{\mathcal{D}_{O}}$, where
$\hat{R}_{d}$ indicates the radius of BH shadow when the observer's
frame lies at the BH position, which is related to $R_{d}$ and
$\mathcal{M}$ is the BH's mass lies at distance $\mathcal{D}_{O}$
from the observer. In this regard, the angular diameter can be
defined as \cite{sd35,sd65}
\begin{equation}\label{20}
D=2\times9.87098\hat{R}_{d}\big(\frac{\mathcal{M}}{M_{\odot}}\big)\big(\frac{1\text{kpc}}{\mathcal{D}_{O}}\big)\mu
as.
\end{equation}
In the case of M$87^{\ast}$, the distance from Earth is
$\mathcal{D}_{O}=16.8\text{kpc}$ and the approximate BH mass is
$\mathcal{M}=(6.5\pm0.7)\times 10^{6}M_{\odot}$, while the actual
shadow diameter is $D_{M87^{\ast}}=(37.8\pm2.7)\mu as$ \cite{sd66}.
Whereas, for Sgr $A^{\ast}$, its distance from Earth is
$\mathcal{D}_{O}=8\text{kpc}$ and its approximated BH mass is
$\mathcal{M}=(4.0^{+1.1}_{-0.6})\times 10^{6}M_{\odot}$, while the
actual shadow diameter is $D_{Sgr A^{\ast}}=(48.7\pm7)\mu as$
\cite{sd67}. We have plotted the shadow angular diameter for
space-time (\ref{s2}) and compared it with EHT results for
M$87^{\ast}$ (top row) and Sgr $A^{\ast}$ (bottom row) in Fig.
\textbf{\ref{israr1}}. For M$87^{\ast}$, the red segments
approximatly lie within $1\sigma$ and $2\sigma$ confidence intervals
under the ranges of $0.1\leq\rho_{c}\leq0.75$ and $0\leq
R_{s}\leq0.3$, see Fig. \textbf{\ref{israr1}} (top row).
Specifically, the lower bound of $\rho_{c}$ and $R_{s}$ is denoted
by $\rho^{min}_{c}=0.4$ and $R^{min}_{s}=0.25$ at which the
transition of shadow diameter is observed from $1\sigma$ to
$2\sigma$ levels. For Sgr $A^{\ast}$, the red segments approximately
lie within $1\sigma$ and $2\sigma$ confidence intervals under the
ranges of $0.1\leq\rho_{c}\leq0.9$ and $0\leq R_{s}\leq0.3$, see
Fig. \textbf{\ref{israr1}} (bottom row). Specifically, the lower
bound of $\rho_{c}$ is denoted by $\rho^{min}_{c}=0.8$ at which the
transition of shadow diameter is observed from $1\sigma$ to
$2\sigma$ levels. While the shadow diameter of Sgr $A^{\ast}$ with
respect to $R_{s}$ lies within $1\sigma$ to $2\sigma$ levels through
out the domain of $R_{s}$. Hence, the Kerr-like BH surrounded by CDM
halo behaves identical with M$87^{\ast}$ and Sgr $A^{\ast}$ under
the allowed range values of $\rho_{c}$ and $R_{s}$, as its
calculated shadow diameter lies within $1\sigma$ to $2\sigma$
levels.

\section{Concluding Remarks}

In this work, we studied an in-depth analysis of the visual
characteristics of rotating BH in the presence of CDM halo, such as
geometrical shape of shadow radius, optical signatures of shadow
images in the background of celestial light sources and a thin
accretion disk, which is consistent with EHT results at $230$GHz.
Initially, we investigated the shadow observed by a distant observer
and analysed how the size of the shadow is influenced by the relevant
parameters of the model. Based on our findings, the spin parameter
$a$ shifts the shadow contours towards the right side, and the space
between the circular orbits is clearer on the left side of the
screen. Further, the increment of critical density $\rho_{c}$ and
the scale radius $R_{s}$ of CDM halo interpret that the shadow
contours are expanded in both cases and have no significant
influence on the deformation of shadow. To analyse the
impact of a strong gravitational field on accreting matter around a
BH, we have investigated the visual characteristics of BHs
illuminated with celestial light sources. The larger values of
$\rho_{c}$ showed that the shadow of D-shape slightly deforms close
to the inner shadow, which is hardly observable. These results are
more clear when both $\rho_{c}$ and $R_{s}$ have larger values.
Moreover, all these images interpret an arc-like shape of Einstein
ring on the left and right sides of the screen. Importantly, the
arc-like shape of the resulting Einstein ring remains stable, while its
position gradually varies with respect to variations of the relevant
parameters.

Based on the thin disk accretion model, we investigated the various
optical characteristics of rotating BH with CDM halo. In this
regard, we analysed the significant features of accreting matter in
which the inner region of accretion disk expands to the BH event
horizon. Moreover, the accreting material is electrically neutral
plasma, and its motion can be classified into two classes based on
ISCO, such as particles move inside and outside the ISCO. For
desired results, we have utilized the fisheye camera model and
conducted a detailed analysis of shadow images of rotating BH, where
the direction of rotation of the accretion disk is classified in two
ways, such as prograde and retrograde accretion flow. In the case of
prgrade accretion flow, we observed that in all images there exists a
central dark region which corresponds to the BH event horizon and a
narrow bright photon ring closely related to the critical curve of
the BH, see Fig. \textbf{\ref{sf5}}. The obtained results indicate
that the intensity as well as the size of the inner shadow are
gradually increased with the aid of $\rho_{c}$ and $R_{s}$. The
clear observation of inner shadow, which shows a hat-like shape,
differentiates the direct and lensed images. Notably, a significant
feature observed on the left side of the screen where a luminous
crescent-like shape appears due to the Doppler effect, enhancing
with the highest values of both $\rho_{c}$ and $R_{s}$. For
an in-depth analysis of accretion material around BH, we have
investigated the red-shift configurations of both direct and lensed
images as depicted in Figs. \textbf{\ref{sf6}} and
\textbf{\ref{sf7}}, respectively. In the context of direct images,
blue and red colour maps appear on the left and right
sides of the screen, respectively. However, a notable difference is
observed that the blue-shift maps are much smaller compared to
the red-shift maps. Moreover, the variations of the parameter space will
result in a slight change to the ranges of red-shift factors, which
are unlikely to prominently impact the total observations. On the
other hand, in the case of lensed images, only the red-shift colour
map appears in the lower right quadrant of the screen. In all
images, the size of the inner shadow is appreciably increased with
the increasing values of relevant parameters.

In Fig. \textbf{\ref{sf8}}, we interpret the lensing bands of
accretion matter under different values of $\rho_{c}$ and $R_{s}$.
The results indicate that with the enlargement of relevant
parameters, the lensed bands (see cyan colour) gradually expand and
significantly deform towards the lower quadrant of the screen.
Moreover, it can be noticed that the position of the photon ring lies
within the range of lensed image, interpreting a smaller amount of
observed flux of lensed emission within the photon ring.
Subsequently, we investigate the image of a rotating BH with CDM halo,
where the direction of accretion flow is retrograde. From Fig.
\textbf{\ref{sf9}}, it can be observed that the intensity of shadow
images is very dim as compared to the prograde accretion flow. Meanwhile,
there are appreciable differences between the images of prograde and
retrograde. Moreover, in this case, a luminous region resembling a
crescent moon emerges on the upper right side of the screen, which
is not detected in the scenario of prograde flow. Moreover, the
red-shift distributions are altered on the screen as compared to
prograde flow. A prominent phenomenon is also observed in Fig.
\textbf{\ref{sf10}}, the red-shift and blue-shift factors are
merged close to the inner shadow, and particularly the red-shift map
expands in more space on the screen as compared to the blue-shift.
Further, the lensed image, in contrast to the prograde flow, moved
towards the lower left side of the screen more obviously, and the
inner shadow region continuously interpreted the hat-like shape with
the increase of critical density $\rho_{c}$. For better analysis, we
also investigated the astrophysical influence of CDM halo on
Kerr-like BH through the comparison of the shadow diameter for
space-time (\ref{s2}) with the recent EHT results of M$87^{\ast}$
and Sgr $A^{\ast}$ as depicted in Fig. \textbf{\ref{israr1}}. The
obtained results confirmed the validity of our considering
space-time under certain parameter constraints. Moreover, the
shadow angular diameter of Sgr $A^{\ast}$ is best fitted with red
segments as compared to M$87^{\ast}$, which provides more parameter
space. Finally, we can conclude that this analysis by saying that the
rotating BH parameters such as the spin parameter $a$, critical density
$\rho_{c}$ and scale radius $R_{s}$ have a prominent impact on the
visual characteristics of the considered space-time. Since in the real
Universe, the astrophysical BHs are expected to be surrounded by
accretion flows by CDM halo. And hence, these findings provide some
new insights into the influence of CDM halo on observation outcomes
and can serve as an important tool for exploring actual nature
of DM.\\\\

{\bf Acknowledgements}\\
This work is supported by the National Natural Science Foundation of
China (Grants No. 11675140, No. 11705005, and No. 12375043),  and
Innovation and Development Joint Foundation of Chongqing Natural
Science  Foundation (Grant No. CSTB2022NSCQ-LZX0021) and Basic
Research Project of Science and Technology Committee of Chongqing
(Grant No. CSTB2023NSCQ-MSX0324).

\end{document}